\documentclass[pdftex,nofootinbib,superscriptaddress,aps,prfluids,notitlepage,longbibliography]{revtex4-2}
\usepackage{multirow,enumerate,epsfig,graphics,amssymb,amsmath,subeqnarray,mathrsfs,color,xcolor,epstopdf}
\usepackage{natbib}
\usepackage{color,epsfig,graphics,amssymb,mathcmd,amsmath,subeqnarray,graphicx,amsthm,subfigure,mathrsfs}
\usepackage[colorlinks=true, citecolor=red, linkcolor=blue ]{hyperref}
\usepackage{mathtools,bm} 
\usepackage{rotating,mathrsfs}
\usepackage[normalem]{ulem}
\usepackage{newtxtext}
\usepackage{newtxmath}
\usepackage{soul}

\makeatletter
\def\sgn{\mathop{\operator@font sgn}}
\def\threevdots{\vbox{\baselineskip1\p@ \lineskiplimit\z@
  \kern6\p@\hbox{.}\hbox{.}\hbox{.}}}
\makeatother

\newcommand{\rb}{\boldsymbol{r}}

\newcommand{\nb}{\boldsymbol{n}}

\begin{document} 

\title{Hydrochemical interactions of phoretic particles: a regularized multipole framework}
\author{Francisco Rojas-Pérez}
\affiliation{LadHyX, CNRS, Ecole Polytechnique, Institut Polytechnique de Paris, 91120 Palaiseau, France}
\affiliation{Departamento de F\'isica, Instituto Tecnol\'ogico de Costa Rica, Cartago, Costa Rica}
\author{Blaise Delmotte}
\email{blaise.delmotte@ladhyx.polytechnique.fr}
\affiliation{LadHyX, CNRS, Ecole Polytechnique, Institut Polytechnique de Paris, 91120 Palaiseau, France}
\author{S\'ebastien Michelin}
\email{sebastien.michelin@ladhyx.polytechnique.fr}
\affiliation{LadHyX, CNRS, Ecole Polytechnique, Institut Polytechnique de Paris, 91120 Palaiseau, France}

\begin{abstract}
Chemically-active colloids modify the concentration of chemical solutes surrounding them in order to self-propel. In doing so, they  generate long-ranged hydrodynamic flows and chemical gradients that modify the trajectories of other particles. As a result, the dynamics of reactive suspensions is fundamentally governed by hydro-chemical interactions. 
A full solution of the detailed hydro-chemical problem with many particles is challenging and computationally expensive. Most current methods rely on the Green's functions of the Laplace and Stokes operators to approximate the particle signatures in the far-field, which is only valid in the very dilute limit in simple geometries.  
To overcome these limitations, we propose a regularized mutipole framework, directly inspired by the Force Coupling Method (FCM), to model phoretic suspensions. Our approach, called Diffusio-phoretic FCM (DFCM), relies on grid-based volume  averages of the concentration field to compute the particle surface concentration moments. These moments define the chemical multipoles  of the diffusion (Laplace) problem and provide the swimming  forcing of the Stokes equations. Unlike far-field models based on singularity superposition, DFCM accounts for mutually-induced dipoles.
The accuracy of the method is evaluated against exact and accurate numerical solutions for a few canonical cases. We also quantify its improvements over far-field approximations for a wide range of inter-particle distances. The resulting framework can readily be implemented into efficient CFD solvers, allowing for large scale simulations of semi-dilute diffusio-phoretic suspensions.
\end{abstract}
\maketitle

\section{Introduction}
Many microscopic organisms and colloidal particles  swim by exerting active stresses on the surrounding fluid in order to overcome its viscous resistance. In doing so, they set their fluid environment into motion and modify the dynamics of their neighbours~\citep*{LaugaPowers2009,ElgetiWrinklerGompper2015}. Large scale collective behaviour can emerge from the resulting long-ranged interactions between individual agents~\citep{PedleyKessler1992,ZottlStark2016}, but also profound modifications of the effective macroscopic rheological and transport properties of such active suspensions~\citep{SaintillanShelley2013,Saintillan2018}. These have recently become a major focus to study a broader class of systems that are fundamentally out of thermodynamic equilibrium, broadly referred to as active matter systems, which comprise large assemblies of individually-active agents that convert locally-stored energy into mechanical actuation  resulting in non-trivial effective macroscopic properties~\citep{MarchettiAllSimha2013,BechingerAllVolpe2016}. 

Most biological swimmers apply such active stresses on the fluid through sequences of shape changes, or swimming strokes, commonly through the flapping of slender flexible appendages such as flagella or cilia~\citep{LaugaPowers2009,BrennenWinet1977,Lauga2016}.  Such cell motility in viscous fluids plays a critical role in a diversity of biological processes including mammal fertility~\citep{FauciDillon2006} 	or the balance of marine life ecosystems~\citep*{GuastoRusconiStocker2012}. Inspired by these biological examples and many promising applications in such various fields as biomedicine or biochemical reactors, researchers and engineers across disciplines have focused on the design of microscopic self-propelled systems~\citep{EbbensHowse2010}. Many earlier designs were directly inspired by the rotation of the helical flagella of bacteria or the flapping of flexible cilia~\citep{DreyfusAllBibette2005,ZhangAllNelson2009,BabataheriAllDuRoure2011}, but rely on complex miniaturization processes of moving parts or a macroscopic actuation (e.g. magnetic fields).

A fundamentally-different route, explored more recently, exploits interfacial processes to generate fluid flow from local physico-chemical gradients (e.g. temperature, chemical potential, electric potential or solute concentration), resulting directly from a chemical activity of the particle surface itself (e.g. catalytic reactions)~\citep{YadavAllSen2015,MoranPosner2017}. The most famous and commonly-used design is that of Janus nano- or micro-particles with two different catalytic or physical properties~\citep{PaxtonAllCrespi2004,PerroAllDuguet2005}. In dilute suspensions, these colloids exhibit short-term ballistic behaviour (with velocities reaching a few $\mu$m.s$^{-1}$) but their long-time dynamics is more diffusive as the result of thermal fluctuations~\citep{HowseAllGolestanian2007}. In contrast, complex collective behaviour is observed in denser suspensions with the coexistence of cluster and gas-like phases~\citep{TheurkauffAllBocquet2012,GinotAllCottinBizonne2018}. Understanding the emergence of such phase-separation is currently a leading challenge in active matter physics~\citep{CatesTailleur2015}. Beyond their fundamental interest and the puzzling details of their individual and collective self-propulsions, these active colloids are already considered for various engineering or biomedical applications, including drug delivery \citep{KaganAllWang2010}, micro-surgery~\citep{ShaoAllvanHest2018}, intelligent cargo delivery~\citep{SundararajanAllSen2008}, self-healing microchips~\citep{LiWang2015}, chemical analysis~\citep{DuanAllSen2015} or sensing~\citep{YiAllYu2016}.

To generate autonomous propulsion, chemically-active colloids exploit a combination of two different physico-chemical properties~\citep*{GolestanianLiverpoolAdjari2007,MoranPosner2017}. The first one is a \emph{phoretic mobility}, namely the ability to generate slip flow along the boundary of a colloidal particle in response to gradients of a solute (diffusiophoresis), temperature (thermophoresis) or electric potential (electrophoresis)~\citep{Anderson1989}, resulting in a net drift of this particle. The second one is the ability of the particle itself to generate the local gradients through a  \emph{surface activity}, e.g. surface-catalysis of chemical reactions~\citep{WangAllMallouk2006} or heat release~\citep{BregullaCichos2015}. The combination of these two generic properties, or \emph{self-phoresis}, provides the colloid with the ability to swim~\citep{GolestanianLiverpoolAdjari2007}. Other self-propulsion mechanisms also share important similarities with self-phoresis, including the propulsion of active droplets~\citep{MaassAllBahr2016} or of light-illuminated colloids in binary mixtures~\citep{ButtinoniAllBechinger2012}. For simplicity,  we focus on self-diffusiophoresis of particles absorbing or releasing neutral chemical solutes { \citep*{Cordova-FigueroaBrady2008,PopescuAllDietrich2016}}, keeping in mind that the approach and framework presented here can be applied or generalised to account for more generic self-phoretic systems {\citep*{MoranPosner2011,Yariv2011,IbrahimAllLiverpool2017}}. 

Symmetry-breaking is an intrinsic requirement for directed motion in viscous flows; for self-phoretic colloids, this requires to create or sustain a chemical surface polarity. As a result, strictly isotropic colloids can not self-propel individually, although they may do so by self-assembling into geometrically- or chemically-asymmetric structures {~\citep{SotoGolestanian2014,SotoGolestanian2015,VarmaMontenegro-JohnsonMichelin2018,SchmidtAllVolpe2019}.} 
 In practice, most chemically-active colloids thus exhibit an intrinsic chemical asymmetry, where the two sides of a Janus colloid capture or release solutes of different natures or at different rates~\citep{MoranPosner2017}. Geometrically-asymmetric colloids also break the symmetry of their chemical environment and may thus self-propel {\citep{KummelAllBechinger2013,ShklyaevAllCordovaFigueroa2014,MichelinLauga2015}}. A third route to symmetry-breaking, based on an instability, arises for isotropic colloids when the chemical solutes diffuse sufficiently slowly for the nonlinear convective coupling of phoretic flows and chemical transport to become significant {~\citep{MichelinLaugaBartolo2013,IzriAllDauchot2014,Hu2019}}.

Like all microswimmers, Janus phoretic particles self-propel by stirring the fluid around them and thus  modify the trajectory and speed of their neighbours. Due to their chemical activity, they also alter their chemical environment and thus also drive an additional phoretic motion of the surrounding particles. {In most experiments on chemically-active particles, the diffusing solutes are small (e.g. dissolved gas) and chemical transport is dominated by diffusion. Such micron-size colloids typically propel with velocities $U\approx$1--10$\mu$m.s$^{-1}$ and consume or release solutes of diffusivity $D\approx 10^3\mu$m$^2$.s$^{-1}$, so that the relevant P\'eclet number $\mbox{Pe}$ is always small ($\mbox{Pe}\approx 10^{-3}$--$10^{-2}$)~\citep{PaxtonAllCrespi2004,HowseAllGolestanian2007,TheurkauffAllBocquet2012,BrownPoon2014}.}Obtaining the swimming velocity of phoretic Janus particles therefore requires solving two different problems sequentially, namely (i) a diffusion (Laplace) problem for the solute concentration around the colloids and (ii) a hydrodynamic (Stokes) problem for the fluid flow around them. Analytical solution is in general amenable only for single particles~\citep{GolestanianLiverpoolAdjari2007}, although determining the coupled motion of two Janus colloids is also possible semi-analytically~\citep{VarmaMichelin2019,NasouriGolestanian2020b,SharifiMood2016}. For more than two particles, a complete description of the phoretic motion requires numerical treatment~\citep*{Montenegro-JohnsonMichelinLauga2015} but with a computational cost that increases rapidly with the number of particles, motivating the use for reduced models for the particles' interactions.

In dilute suspensions, i.e. when particles are far apart from each other, their hydro-chemical interactions can be accounted for through the slowest-decaying chemical and hydrodynamic signatures of individual particles and their effect on their neighbours~\citep{SahaGolestanianRamaswamy2014,VarmaMichelin2019}.  Due to their simplicity, small computational cost for large number of particles and  ability to handle the effect of confinements through image systems, far-field models have been extensively used to analyse the motion of active suspensions{~\citep[see e.g.][]{IbrahimLiverpool2016,Thutupalli2018,KansoMichelin2019,LiebchenLowen2019}}. An alternative mean-field approach describes the particles' motion in the ambient chemical and hydrodynamic fields generated by the superposition of their individual far-field signatures~\citep{LiebchenAllCates2015,TraversoMichelin2020}.

For more concentrated suspensions, i.e. when the inter-particle distance is reduced, far-field models are not accurate as finite-size effects of the particles are no longer negligible. Although it is possible to include higher order corrections using the Method of Reflections~\citep{VarmaMichelin2019}, more complex numerical models are in general required to solve the dual hydro-chemical problem accurately within not-so-dilute suspensions. Due to the mathematical similarities between Laplace and Stokes problems, it is possible to draw inspiration from and build upon a large variety of methods already used in recent years for the numerical modelling of passive and active suspensions. A popular example is the Stokesian dynamics and its more recent extensions~\citep{BradyBossis1988,SwamBradyMoore2011,SierouBrady2001,FioreSwan2019}, from which an analogous approach was proposed to solve for diffusion problems~\citep{YanBrady2016}.  {A similar approach relies on a truncated spectral expansion of the integral formulation of the Laplace and Stokes equations with tensorial spherical harmonics on the particle's surface~\citep{Singh2019,Singh2019pystokes}}. But the possible routes also include Boundary Element Methods~\citep{IshikawaSimmondsPedley2006,UspalAllTasinkevych2015,Montenegro-JohnsonMichelinLauga2015},   Immersed Boundary Methods~\citep{LushiPeskin2013,LambertAllBrandt2013,BhallaAllDonev2013}, Lattice-Boltzmann approaches~\citep{AlarconPagonabarraga2013,LaddVerberg2001}, Multi-Particle Collision Dynamics~\citep*{ZottlStark2014,YangWysockiRipoll2014,ColbergKapral2017,ZottlStark2018},
and the Force Coupling Method~\citep{MaxeyPatel2001,DelmotteAllCliment2015}. 

The objective of the present work is to extend the fundamental idea and framework of the latter {to establish and validate a unified method that accounts for both  chemical and hydrodynamic interactions between phoretic particles.} The Force Coupling Method (FCM) used to solve for the hydrodynamic interactions of particles in a fluid relies on the classical multipolar expansion of the solution for Stokes' equation~\citep{Saffman1973}, but proposes a regularised alternative to singular Green's function in the form of smoothed Gaussian kernels. Beyond the obvious numerical advantage of such a regularization, it also provides an indirect route to account for the finite size of the particles through the finite support of these kernels. The FCM framework was initially proposed twenty years ago by Maxey and coworkers~\citep{MaxeyPatel2001,LomholtMaxey2003} to analyse the joint dynamics of passive spherical particles sedimenting in a viscous fluid. It has since then been extended to account for finite inertia~\citep*{XuMaxeyKarniadakis2002}, lubrication effects~\citep{DanceMaxey2003} and non-sphericity of the particles~\citep{LiuAllKarniadakis2009} leading to a powerful method to study the hydrodynamic interactions of large suspensions. More recently, FCM was also adapted to account for the activity of the colloids and enabled the analysis of microswimmer suspensions~\citep{DelmotteAllCliment2015}.

 In this work, an FCM-based method is presented to solve the Laplace problem for the concentration field in phoretic suspensions of spherical Janus particles, using a regularized multipole representation of the concentration based on smoothed kernels instead of the classical singular monopole and dipole singularities. This provides the phoretic forcing introduced by the local inhomogeneity of the concentration field on each particle, from which  the hydrodynamic problem can be solved using the existing FCM approach for active suspensions~\citep{DelmotteAllCliment2015}. Taken together, this provides an integrated framework to solve for the complete diffusiophoretic problem, or Diffusiophoretic Force Coupling Method whose fundamental justification and validation is the main objective of the present work.

The rest of the paper is organized as follows. The governing equations for the collective motion of phoretic particles are first reminded in Section~\ref{sec:equations}. The Diffusiophoretic Force Coupling Method (DFCM) is then presented in detail in Section~\ref{sec:dfcm}. More specifically, the new solution framework for the Laplace problem is first presented in Section~\ref{sec:ReactiveFCM}. Section~\ref{sec:hydroFCM} summarizes the main elements of the classical hydrodynamic FCM method and its extension to active particles, and Section~\ref{sec:coupling} finally presents how the two steps are conveniently coupled to solve successively the chemical and hydrodynamic {problems}. In order to validate the approach and compare its accuracy to existing methods, Section~\ref{sec:Results} considers a series of canonical configurations for pairwise interactions of two Janus particles, for which an analytical or numerical solution of the full problem is available for any inter-particle distance. The results of DFCM are compared to this benchmark but also to the far-field estimation of the particles' velocities. This provides further insight on the improvement brought by this approach and its range of validity, which will be a critical information for future use in larger suspension simulations. Finally, Section~\ref{sec:conclusions} summarizes the findings of the paper, the constraints and advantages of the  method and discusses some perspectives for its future implementation in studying large phoretic suspensions.

\section{Modelling reactive suspensions}\label{sec:equations}
Reactive suspensions consist of large sets of  micro-particles that are able to self-propel in a viscous fluid by exploiting the chemical activity of their surface and its ability to generate an effective hydrodynamic slip in response to gradients of the solute species they produce or consume. As a result, these particles react to the chemical and hydrodynamic forcing exerted by their neighbours, introducing a coupling that may lead to modified effective properties at the scale of the suspensions. For purely diffusive solute species, determining their individual dynamics requires solving successively for two different problems, namely a Laplace problem for the solute concentration distribution, followed by a Stokes problem for the hydrodynamic fields and particle velocities (translation and rotation) in response to the solute gradients at their surface~\citep{GolestanianLiverpoolAdjari2007}. The corresponding equations of motion are reminded in detail below.

\subsection{Governing equations for self-diffusiophoresis of $N$ micro-particles}
\begin{figure}
    \begin{center}
    \raisebox{2.4in}{\small a)}\includegraphics[width=0.50\textwidth]{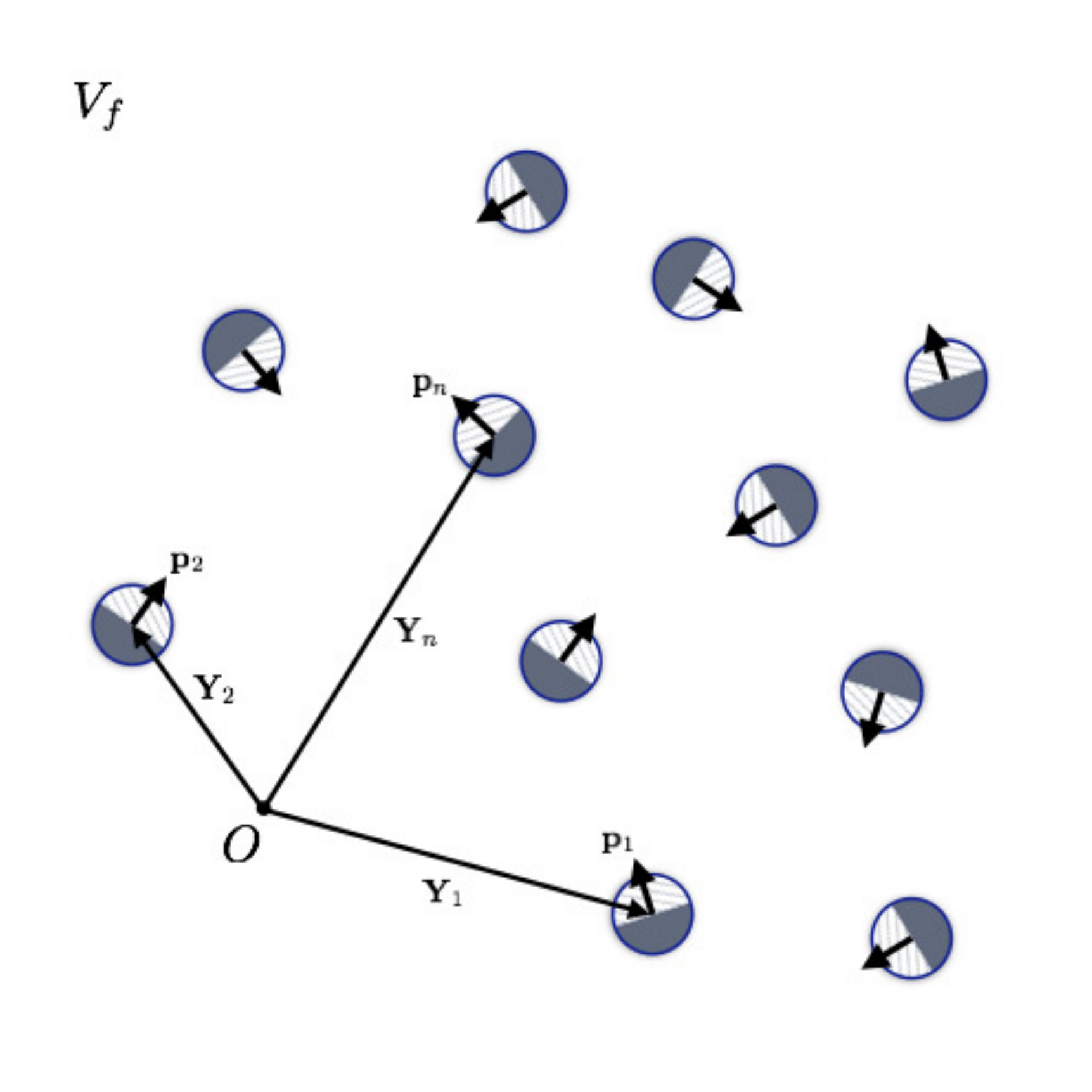} \quad \quad \quad
    \raisebox{1.8in}{\small b)} \quad \includegraphics[width=0.22\textwidth]{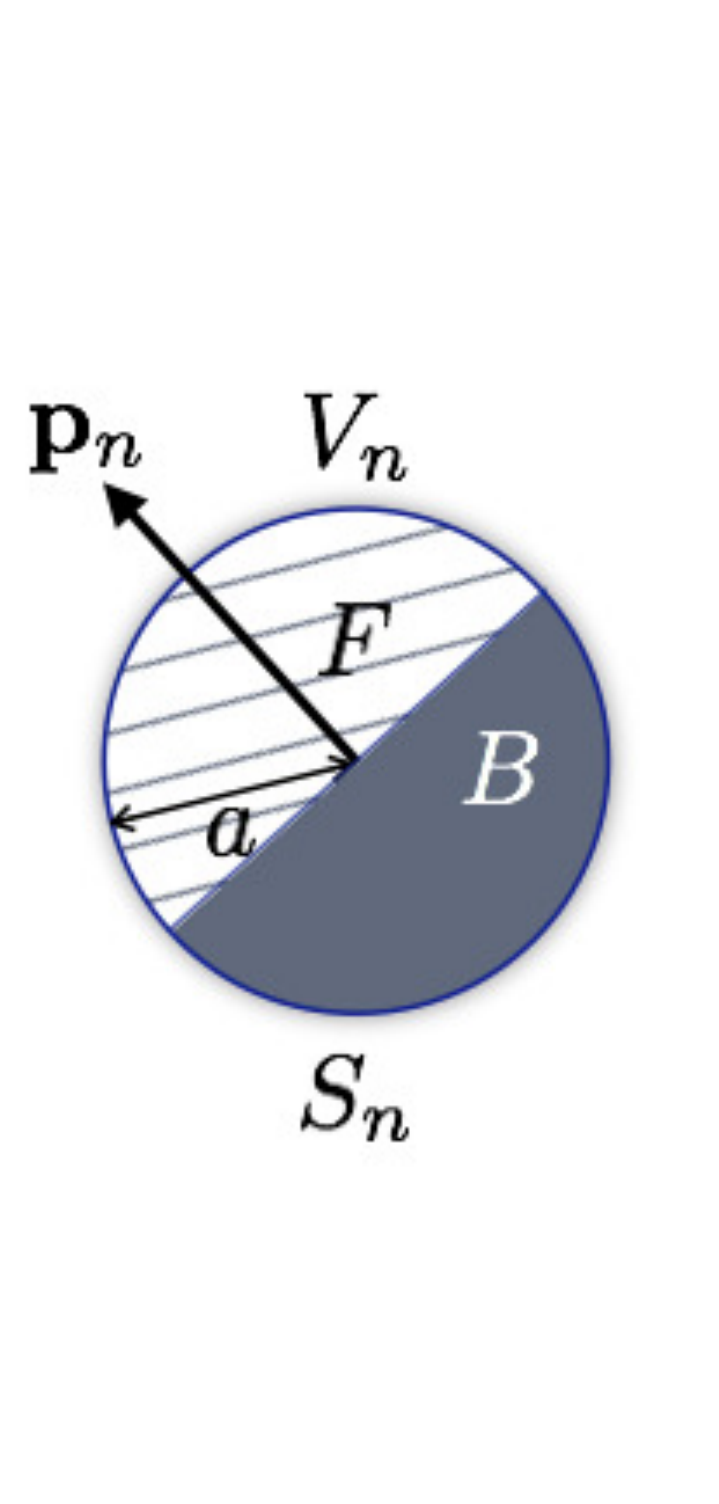}
    \end{center}
    \caption{(a) Geometric description and parameter definition for (a) a reactive suspension system and (b) an individual active particle including the fluid domain $V_f$, as well the phoretic particles' position $\boldsymbol{Y}_n$ and orientation $\boldsymbol{p}_n$, their radius $a$. The particle's orientation $\boldsymbol{p}_n$, allows for the definition of its front caps (noted $F$ and $B$ respectively). The different colours of the caps (white or grey) illustrate their different chemical activity, while their pattern (striped and solid) illustrate their different mobilities.}
    \label{fig:suspensionSystemGeometry}
\end{figure}

The coupled motion of $N$ identical and spherical phoretic particles of equal radius $a$ is considered within a viscous fluid of density $\rho$ and viscosity $\mu$. Particle $n$ occupies a volume $V_n$ bounded by its surface $S_n$ and centred at $\boldsymbol{Y}_n(t)$, and has orientation $\boldsymbol{p}_n$; $\boldsymbol{U}_n$ and  $\boldsymbol{\Omega}_n$ are its translation and rotation velocities. The fluid domain is noted $V_f$ and may be bounded or unbounded (figure \ref{fig:suspensionSystemGeometry}a).

Each particle emits  a chemical solute of diffusivity $D$ on the catalytic parts of its surface with a fixed spatially-dependent rate, of characteristic magnitude $\alpha_0$, and is able to generate a slip flow in response to a surface concentration gradient, with a characteristic phoretic mobility $M_0$. In the following, all variables and equations are made dimensionless using $a$, $U_0=\alpha_0 M_0/D$ and $a\alpha_0/D$ as characteristic length, velocity and concentration scales. 

As a result of its surface activity, the dimensionless relative concentration $c$ (with respect to its background value far from the particles) satisfies the following Neumann condition on the surface of particle $n$: 
\begin{equation}
    -\boldsymbol{n} \cdot \nabla c = \alpha_n(\mathbf{n}) \quad \quad \mathrm{on} \ S_n,
    \label{eq:boundaryConditionDiffusionSphereN}
\end{equation}
 where $\alpha_n(\boldsymbol{n})$ is the dimensionless activity distribution (i.e. emission rate) and $\boldsymbol{n}$ is the outward normal unit vector on $S_n$.
For sufficiently small particles, the solute's dynamic is purely diffusive, i.e. the relevant P\'eclet number ${\mbox{Pe}=aU_0/D\ll 1}$, so that $c$ obeys Laplace's equation outside the particles,
\begin{equation}
    \nabla^2 c = 0 \quad \quad \mathrm{in} \ V_f.
    \label{eq:LaplaceEquation}
\end{equation}
Together with an appropriate boundary conditions at the external boundary of $V_f$ (e.g. $c\rightarrow 0$ for $|\boldsymbol{r}|\rightarrow \infty$ in unbounded domains), these equations form a well-posed problem for the distribution of solute in the fluid domain $V_f$.

In response to non-uniform solute distribution at the particles' surface, a phoretic slip flow $\boldsymbol{u}^s_n$ develops outside a thin interaction layer {\citep{Anderson1989}} so that effectively, the hydrodynamic boundary condition on $S_n$ becomes 
\begin{equation}
    \boldsymbol{u} = \boldsymbol{U}_n + \boldsymbol{\Omega}_n \times \boldsymbol{r}_n + \boldsymbol{u}^s_n,\qquad \textrm{with   }\boldsymbol{u}_n^s=M_n(\mathbf{n}) \nabla_{||} c\quad \quad \mathrm{on} \ S_n.
    \label{eq:boundaryConditionHydrodynamicsSphereN}
\end{equation}
In the previous equation, $\nabla_{||}=(\mathbf{I}-\boldsymbol{n}\boldsymbol{n}) \cdot \nabla$ is the tangential gradient on the particle's surface, $\boldsymbol{r}_n=\boldsymbol{r}-\boldsymbol{Y}_n$ is the generic position relative to {particle $n$'s} centre, and  $M_n(\boldsymbol{n})$ denotes the dimensionless and spatially-dependent phoretic mobility of the surface of particle $n$. For small particles, inertial effects are negligible (i.e $\mbox{Re} = \rho U_0 a/\mu \ll 1$), and the dimensionless fluid's velocity and pressure ($\boldsymbol{u}, p$) satisfy  Stokes' equations:
\begin{equation}
    \nabla p = \nabla^2 \boldsymbol{u},\qquad \nabla\cdot \boldsymbol{u}=0 \quad \quad \mathrm{in} \ V_f,
    \label{eq:StokesEquation}
\end{equation}
with appropriate condition at the outer boundary of $V_f$ (e.g. $\boldsymbol{u}\rightarrow 0$ for $|\boldsymbol{r}|\rightarrow\infty$). Neglecting any outer forcing such as gravity, each particle is hydrodynamically force- and torque-free {\citep{PopescuAllDietrich2016}} at all times,
\begin{equation}
    \boldsymbol{F}_n = \int_{S_n} \boldsymbol{\sigma} \cdot \boldsymbol{n} \ \mathrm{d}S = \boldsymbol{0}, \qquad \qquad
    \boldsymbol{T}_n = \int_{S_n} \boldsymbol{r}_n \times (\boldsymbol{\sigma} \cdot \boldsymbol{n}) \ \mathrm{d}S = \boldsymbol{0},
    \label{eq:forceFreeTorqueFreeCondition}
\end{equation}
with $\boldsymbol\sigma=-p\mathbf{I}+(\nabla\mathbf{u}+\nabla\mathbf{u}^T)$ the dimensionless Newtonian stress tensor, and their dominant hydrodynamic signature is therefore that of a force dipole or stresslet $\mathbf{S}_n$~\citep{Batchelor1970}.

For a given concentration distribution $c$, Equations~\eqref{eq:boundaryConditionHydrodynamicsSphereN}--\eqref{eq:forceFreeTorqueFreeCondition} form a well-posed problem for the fluid velocity and pressure, and particle velocities, so that at a given time $t$, and for given particle positions and orientations, $\boldsymbol{Y}_n(t)$ and $\boldsymbol{p}_n(t)$, the successive Laplace and Stokes problems presented above uniquely determine the instantaneous particle velocities $\boldsymbol{U}_n(t)$ and $\boldsymbol{\Omega}_n(t)$, from which the motion of the particles is obtained:
\begin{equation}
    \frac{\mathrm{d}\boldsymbol{Y}_n}{\mathrm{d}t} = \boldsymbol{U}_n,\qquad 
    \frac{\mathrm{d}\boldsymbol{p}_n}{\mathrm{d}t} = \boldsymbol{\Omega}_n \times \boldsymbol{p}_n.
    \label{eq:odeForParticleOrientationN}
\end{equation}

For a single isolated particle, the Lorentz Reciprocal Theorem to Stokes flows provides the particle's translation and rotation velocities directly in terms of the phoretic slip~\citep{StoneSamuel1996}:
\begin{equation}
    \boldsymbol{U} = - \langle \boldsymbol{u}^s \rangle,    \qquad \qquad      \boldsymbol{\Omega} = -\frac{3}{2a}\langle \boldsymbol{n} \times \boldsymbol{u}^s \rangle,
    \label{eq:translationalAndRotationalVelocitiesCoupling}
\end{equation}
where $\langle \cdot \rangle$ is the spatial average  over the particle's surface. Similarly, the stresslet $\mathbf{S}$ of the particle is obtained as~\citep{LaugaMichelin2016}, 
\begin{equation}
    \mathbf{S} = -10\pi a^2 \langle \boldsymbol{n} \boldsymbol{u}^s +\boldsymbol{u}^s \boldsymbol{n} \rangle.
    \label{eq:stressletCoupling}
\end{equation}

\subsection{Hemispheric Janus phoretic particles}
Most phoretic particles have a Janus-type surface consisting of two different materials or surface coatings with distinct physico-chemical properties (e.g. a catalytic side and a passive one)~ \citep{PaxtonAllCrespi2004,HowseAllGolestanian2007,TheurkauffAllBocquet2012}. These provide the particles with a built-in chemical asymmetry that triggers the inhomogeneity of the concentration distribution at their surface at the heart of their self-propulsion. In the following, we thus consider such hemispheric Janus particles with uniform but distinct mobilities $(M_n^F,M_n^B)$ and activities $(\alpha^F_n,\alpha_n^B)$ on their front (F) and back (B) hemispheres, as defined with respect to their orientation $\boldsymbol{p}_n$ (figure \ref{fig:suspensionSystemGeometry}b), e.g.  the surface mobility of particle $n$ writes 
\begin{equation}
    M_n(\boldsymbol{n}) = \overline{M}_n + M^*_n \ \mathrm{sign}(\boldsymbol{p}_n \cdot \boldsymbol{n}),
    \label{eq:trueJanus_1_2_mobiAlteExpr}
\end{equation}
with $\overline{M}_n=(M_n^F+M_n^B)/2$ and $M_n^*=(M_n^F-M_n^B)/2$ the mean mobility and mobility contrast, and a similar definition for the spatially-dependent activity $\alpha_n(\boldsymbol{n})$ at the particle's surface. The special case of a particle with uniform mobility thus corresponds to $\overline{M}_n=M^0_n$ and $M^*_n=0$.

\section{An FCM-based method for phoretic suspensions} \label{sec:dfcm}

In the purely diffusive and viscous limit, solving for the particles' dynamics therefore amounts to solving sequentially two linear problems, namely a Laplace problem for $c$ and a Stokes swimming problem for the hydrodynamic field and particles' velocity. Although the exact solution to this joint problem can be obtained analytically for the single- and two-particle cases~\citep{GolestanianLiverpoolAdjari2007,SharifiMood2016,VarmaMichelin2019}, analytical treatment becomes intractable beyond $N\geq 3$ due to the geometric complexity of the fluid domain and despite the problem's linearity. Numerical simulations are therefore critically needed, and several numerical strategies have been proposed recently and briefly reviewed in the introduction. In order to analyse accurately the collective dynamics of in a suspension of Janus phoretic particles, such a method must combine an efficient solution of the Laplace and Stokes problems outside a large number of finite-size objects, while providing accurate representation of the coupling at the surface of each particle between chemical and hydrodynamic fields.

With that double objective in mind, we propose and present here a novel numerical framework to solve  for the reactive suspension problem presented in Section~\ref{sec:equations}, based on the classical Force Coupling Method (FCM) used for pure hydrodynamic simulations of passive particles or microswimmers, thereby generalising its application to the solution of the chemical diffusion problem and its coupling with the already-established hydrodynamic FCM~\citep{MaxeyPatel2001,LomholtMaxey2003,YeoMaxey2010, DelmotteAllCliment2015}. Section~\ref{sec:ReactiveFCM} develops the regularized Laplace problem and associated Reactive FCM, while Sec.~\ref{sec:hydroFCM} presents a brief review of the existing hydrodynamic FCM, and Sec.~\ref{sec:coupling} combines both to obtain a new Diffusio-phoretic Force Coupling Method approach.

\begin{figure}
    \begin{center}
    \raisebox{2.4in}{\small a)}\includegraphics[width=0.50\textwidth]{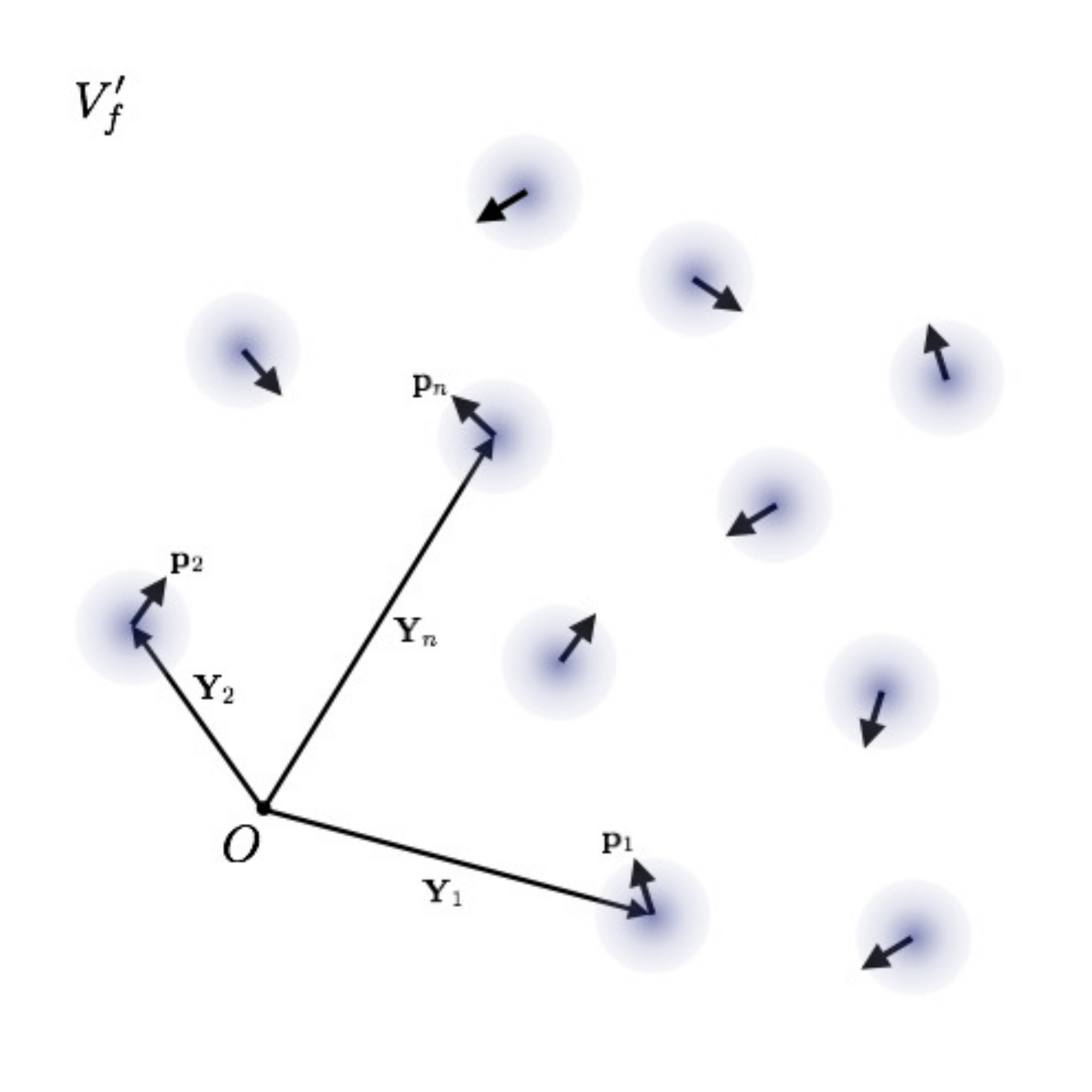}\quad \quad \quad
    \raisebox{1.8in}{\small b)} \quad \includegraphics[width=0.22\textwidth]{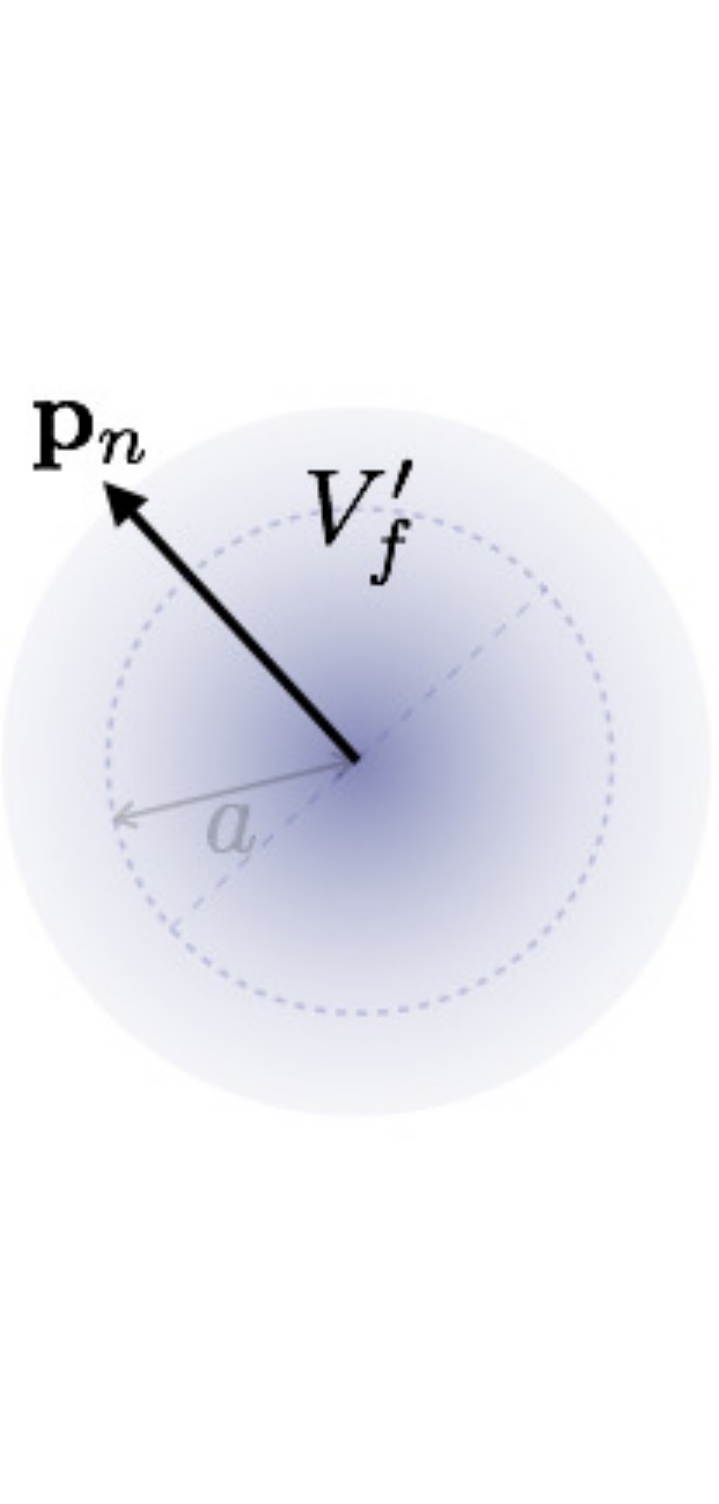}
    \end{center}
    \caption{Regularized representation of (a) the reactive suspension system and (b) individual particles in the DFCM framework. The chemical and hydrodynamic fields are now defined over the entire domain with distributed forcings defined relative to each particle's position $\boldsymbol{Y}_n$ and orientation $\boldsymbol{p}_n$. The boundary $S_n$ of the real particle (dashed) and its radius $a$ are plotted only as reference.}
    \label{fig:approximateSuspensionSystemGeometry}
\end{figure}

The fundamental idea of the Force Coupling Method is to replace a solution of the Stokes equations only within the fluid domain $V_f$ \emph{outside} the forcing particles, by a solution of these equations over the entire domain ${V_F=V_f\cup V_1\cup\hdots\cup V_N}$ (i.e. both outside and inside the particles), replacing the surface boundary conditions with a distributed regularised forcing over a compact envelope calibrated so as to reproduce certain physical features of the problem and account for a weak form of the surface boundary conditions (figure~\ref{fig:approximateSuspensionSystemGeometry}). Doing so, the costly discrete resolution and time-dependent meshing of the particles is no longer necessary, so that efficient (e.g. spectral) Laplace and Stokes solvers on a fixed regular grid may be used at all times, offering significant performance and scalability advantages with respect to other approaches (e.g. Boundary Element Methods). More specifically, FCM associates to each particle a finite set of regularized hydrodynamic singularities (force monopoles, dipoles and so on) chosen so as to satisfy a weak form of the surface boundary conditions.

\subsection{Reactive FCM}\label{sec:ReactiveFCM}
We extend here this approach to the solution of the Laplace problem for $c$ in Eqs.~\eqref{eq:boundaryConditionDiffusionSphereN}--\eqref{eq:LaplaceEquation}. Replacing each particle by a distributed forcing modifies Laplace's equations into a Poisson equation over the entire domain  $V_F$ (including both fluid and particles), 
\begin{equation}
    \nabla^2 c = -g(\boldsymbol{r},t) \quad \quad \mathrm{in} \ V_F,
    \label{eq:LaplaceEquationModified}
\end{equation}
where the function $g(\boldsymbol{r},t)$ includes the source terms accounting for the presence of each particle.

\subsubsection{Standard Multipole Expansion for Laplace problem}
\begin{figure}
    \begin{center}
    \raisebox{1.7in}{\small a)}\includegraphics[width=0.45\textwidth]{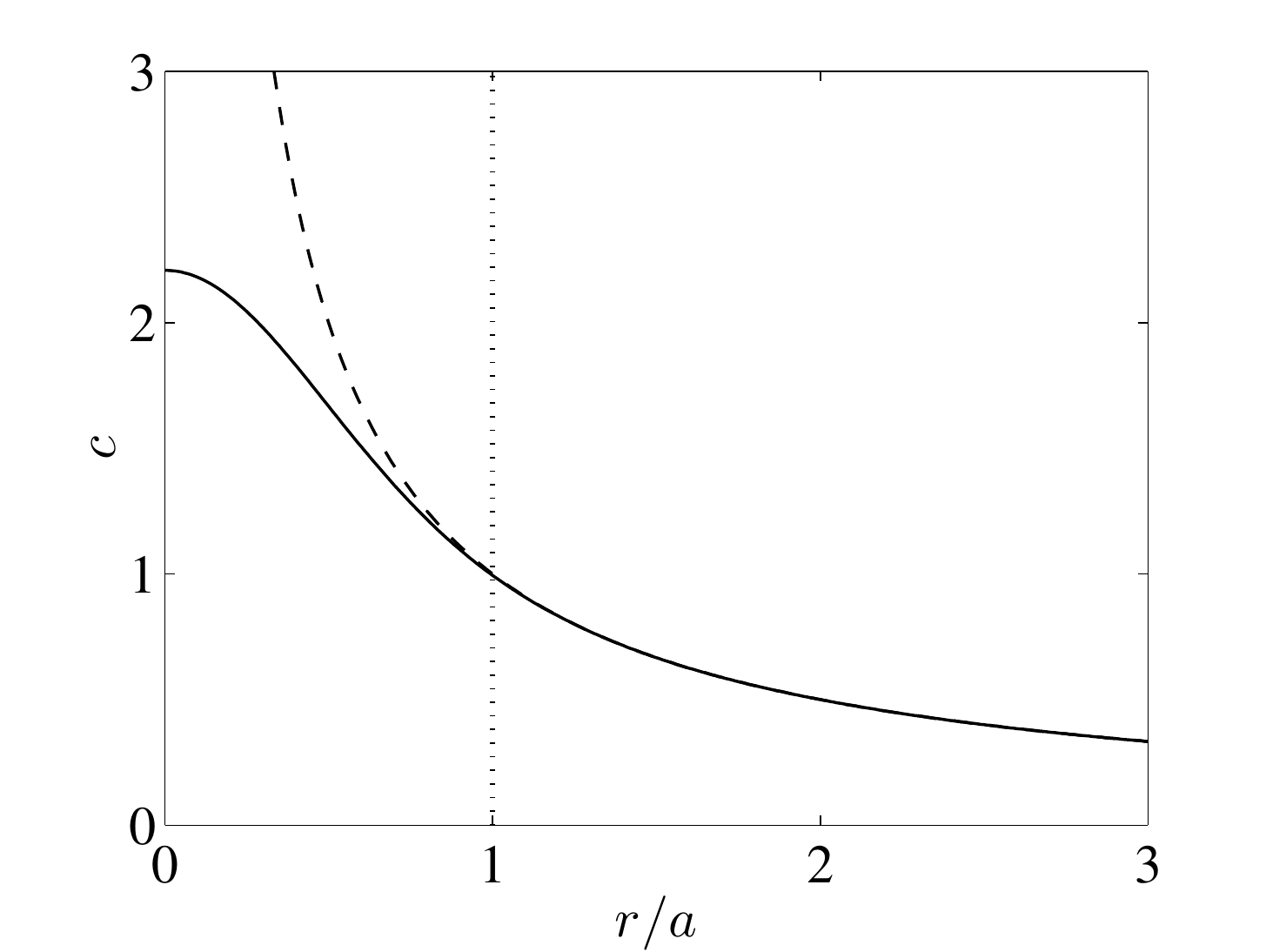}\quad
    \raisebox{1.7in}{\small b)}\includegraphics[width=0.45\textwidth]{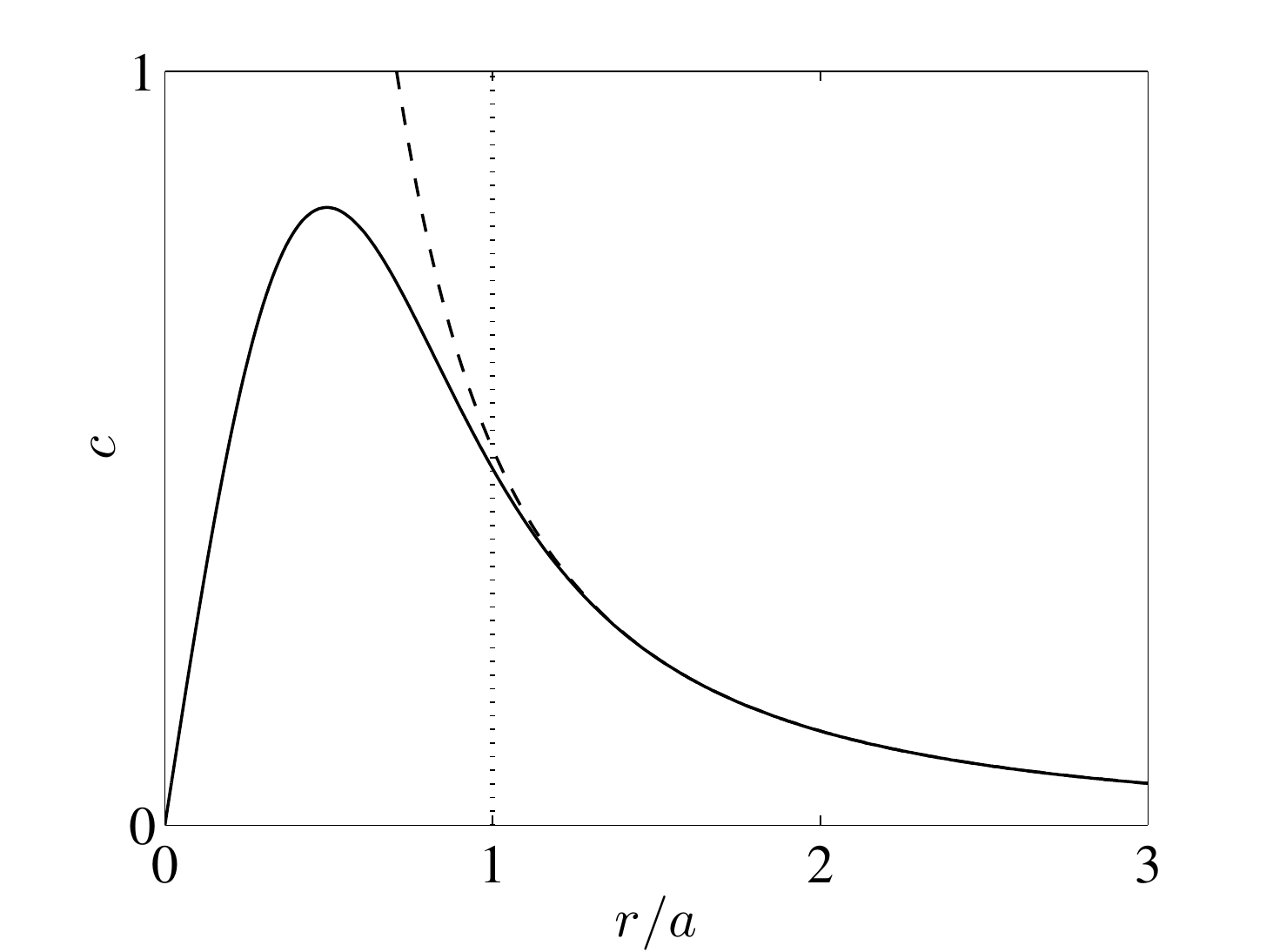}
    \end{center}
    \caption{Singular (dotted lines, Eq.~\eqref{eq:MonopoleDipoleSingularSolution}) and regularized (solid lines, Eqs.~\eqref{eq:MonopoleregularizedSolution}--\eqref{eq:DipoleregularizedSolution}) concentration distributions along the axial polar direction associated to the Greens' Functions for the Laplace equation for: a) monopole terms and b) dipole terms. The line $r/a=1$ represents the particle surface.}
    \label{fig:greensFunctionsFCMandSME}
\end{figure}

The exact solution of the Laplace problems can in fact be recovered from Eq.~\eqref{eq:LaplaceEquationModified}, when the function $g(\boldsymbol{r},t)$ is taken as a (possibly infinite) set of singularities centred on each particle~\citep{Saffman1973},
\begin{equation}
    g(\boldsymbol{r},t) = \sum_{n=1}^N \Big[ q^M_n \delta(\boldsymbol{r}_n) {-} \boldsymbol{q}^D_n \cdot \nabla \delta(\boldsymbol{r}_n) + ...\Big],
    \label{eq:LaplaceEquationModified_RHS_SME}
\end{equation}
where $\delta(\boldsymbol{r}_n)$ is the Dirac delta distribution, and ($q_n^M$, $\boldsymbol{q}_n^D$,...) are the {intensities} of the singularities associated with particle $n$, and are constant tensors of increasing order. {Note that $\nabla$ denotes here the gradient with respect to the observation position $\boldsymbol{r}$ and $\boldsymbol{r}_n=\boldsymbol{r}-\mathbf{Y}_n$.} This equation can be solved explicitly for the concentration field $c$ as a multipole expansion  for each particle in terms of source monopoles, dipoles, etc...
\begin{equation}
    c(\boldsymbol{r},t) = \sum_{n=1}^N \Big[ q^M_n G^M(\boldsymbol{r}_n) + \boldsymbol{q}^D_n \cdot \boldsymbol{G}^D(\boldsymbol{r}_n) + ...\Big],
    \label{eq:cFieldMultipoleExpansion}
\end{equation}
where $G^M$ and $\boldsymbol{G}^D$ are the monopole and dipole Green's functions and satisfy 

\begin{equation}
    \nabla^2 G^M= -\delta(\boldsymbol{r}_n),\qquad \nabla^2 \boldsymbol{G}^D= \nabla\delta(\boldsymbol{r}_n),
    \label{eq:LaplaceEquationModifiedForMonopoleNsingular}
\end{equation}
together with appropriate decay or boundary conditions on the domain's outer boundary. For unbounded domains with decaying conditions in the far-field, the singular monopole and dipole Green's functions are simply
\begin{equation}
    G^M(\boldsymbol{r}_n) = \frac{1}{4\pi r_n}\quad \textrm{and }\quad\boldsymbol{G}^D(\boldsymbol{r}_n)=-\nabla G^M =\frac{\boldsymbol{r}_n}{4\pi r_n^3}\cdot
    \label{eq:MonopoleDipoleSingularSolution}
\end{equation}
The concentration distributions associated to these singular Green's functions are displayed in figure \ref{fig:greensFunctionsFCMandSME}. Higher-order derivatives of $G^M(\boldsymbol{r})$, Eq.~\eqref{eq:MonopoleDipoleSingularSolution}, are also solutions of Laplace's equation leading to singularities of increasing order (quadrupole, octopole,...).

\subsubsection{Truncated regularized multipole expansion}

The previous approach, based on an infinite set of singular sources, is known as the standard multipole expansion of the Laplace problem. Although satisfying from a theoretical point of view, since it is able to recover an accurate representation of the analytical solution outside the particles for a large enough number of singular {multipoles}, it is not well-suited for a versatile numerical implementation because of (i) the singular behaviour of the forcing terms in the modified Laplace equation, Eq.~\eqref{eq:LaplaceEquationModified}, and (ii) the \emph{a priori} infinite set of singularities required for each particle.

To avoid the latter issue, the infinite expansion is truncated here after the first two terms, thus retaining the monopole and dipole contributions only. Physically, this amounts to retaining the two leading physical effects of the particle on the concentration field, i.e. a net emission with a front-back asymmetric distribution. In order to overcome the former problem, the standard FCM replaces the singular Dirac distributions $\delta(\boldsymbol{r})$ by regular Gaussian spreading functions $\Delta(\boldsymbol{r})$:
\begin{equation}
    \triangle(\boldsymbol{r}) = (2\pi\sigma^2)^{-3/2} \mathrm{exp}\Big(-\frac{r^2}{2\sigma^2}\Big),
    \label{eq:regularizedSpreadingEnvelopes}
\end{equation}
where $\sigma$ denotes the finite-size support of this envelop and acts as a smoothing parameter of the method, thus eliminating the singular behaviour of the delta distribution $\delta(\boldsymbol{r})$ near the origin, thereby allowing for a more accurate numerical treatment. The original singular distribution is recovered when $\sigma \ll r$, i.e. the solution of the regularised problem is an accurate representation of the true solution away from the particle. 
This approach using regular distributions allows for a more versatile and robust numerical solution of the physical equations than their singular counterparts~\citep{MaxeyPatel2001,LomholtMaxey2003}.

Combining these two approximations, we therefore consider a truncated regularized expansion including only the monopole and the dipole terms as:
\begin{equation}
    g(\boldsymbol{r},t) = \sum_{n=1}^N \Big[ q^M_n \Delta^M({r}_n) {-} \boldsymbol{q}^D_n \cdot \nabla \Delta^D({r}_n)\Big],
    \label{eq:LaplaceEquationModified_RHS_RME}
\end{equation}
with the Gaussian spreading operators $\Delta^M$ and $\Delta^D$ defined as:
\begin{equation}
    \Delta^M(r) = (2\pi\sigma_M^2)^{-3/2} \mathrm{exp}\Big(-\frac{r^2}{2 \sigma_M^2}\Big), \qquad
    \Delta^D(r)  = (2\pi\sigma_D^2)^{-3/2} \mathrm{exp}\Big(-\frac{r^2}{2 \sigma_D^2}\Big),
    \label{eq:envelopeMonopoleDipoleChemistry}
\end{equation}
where $M$ and $D$ once again denotes monopole and dipole, and $\sigma_M$ and $\sigma_D$ are the finite support of each regularized distribution and are free numerical parameters of the method that need to be calibrated. Note that in all generality, these do not need to be identical~\citep{LomholtMaxey2003}.

 The corresponding truncated regularized solution for $c$ is then finally obtained as:
\begin{equation}
    c(\boldsymbol{r},t) = \sum_{n=1}^N \Big[ q^M_n G^M(\rb_n) + \boldsymbol{q}^D_n \cdot \boldsymbol{G}^D(\rb_n) \Big],
    \label{eq:cFieldTruncatedMultipoleExpansion}
\end{equation}
with the regularized monopole and dipole Green's functions
\begin{align}
    G^M(\rb) &= \frac{1}{4\pi r} \mathrm{erf}\Big( \frac{r}{{\sigma_M}\sqrt{2}} \Big), \label{eq:MonopoleregularizedSolution}\\
     \boldsymbol{G}^D(\boldsymbol{r})&= \frac{\boldsymbol{r}}{4\pi r^3} \Big[ \mathrm{erf}\Big(\frac{r}{\sigma_D \sqrt{2}}\Big) -\sqrt{\frac{2}{\pi}} \Big(\frac{r}{\sigma_D}\Big) \mathrm{exp}\Big(-\frac{r^2}{2\sigma_D^2}\Big) \Big].
    \label{eq:DipoleregularizedSolution}
\end{align}
These clearly match the behaviour of their singular counterpart, Eq.~\eqref{eq:MonopoleDipoleSingularSolution}, when $r$ is greater than a few $\sigma_M$ or $\sigma_D$, respectively, while still maintaining finite values within the particle (figure~\ref{fig:greensFunctionsFCMandSME}), e.g. $\boldsymbol{G}^D(\boldsymbol{r}=\boldsymbol{0})=\boldsymbol{0}$.

\subsubsection{Finding the intensity of the singularities}
Up to this point, no information was implemented regarding the surface boundary conditions on $c$ in Eq.~\eqref{eq:boundaryConditionDiffusionSphereN}. We now present how to determine the intensities of the monopole and dipole distributions associated with each particle, $q^M_n$ and $\boldsymbol{q}^D_n$,  so as to satisfy a weak form of the Neuman boundary condition, Eq.~\eqref{eq:boundaryConditionDiffusionSphereN}, i.e. its first two moments over the particle's surface.  Using the multipole expansion of the fundamental integral representation of the concentration (see Appendix~\ref{app:intensities}), the monopole and dipole intensities of particle $n$, $q^M_n$ and $\boldsymbol{q}^{D}_n$, are obtained as~\citep{YanBrady2016}: 
\begin{equation}
    q^M_n = \int_{S_n} \alpha_n \mathrm{d}S,\qquad \boldsymbol{q}^{D}_n = a \int_{S_n} \alpha_n \boldsymbol{n} \mathrm{d}S + 4\pi a^2 \langle c \boldsymbol{n} \rangle_n
    \label{eq:monopoledipoleIntensity_FINAL}
\end{equation}
where the second term in $\boldsymbol{q}_n^D$ is proportional to the concentration polarity at the surface of particle $n$, i.e. its first moment $\langle c \boldsymbol{n} \rangle_n$,  and is defined using the surface average operator $\langle\cdot\rangle_n$ over particle $n$'s surface.
Note that the activity distribution at the particle's surface is known, and thus Eq.~\eqref{eq:trueJanus_1_2_mobiAlteExpr} explicitly provides the monopole intensity and the first term in the dipole intensity. The second contribution to the latter requires however knowledge of the solution on the particle's surface -- which is not explicitly represented in the present FCM approach. This term therefore requires to be solved for as part of the general problem. In the previous equation, it should be noted that the dimensionless particle radius is $a=1$, but will be kept in the equations to emphasize the relative scaling of the numerical spreading enveloppes (e.g. $\sigma_M$ and $\sigma_D$) with respect to the particle size.

 Here, we use an iterative approach to solve this linear joint problem for the dipole intensity and concentration field, solving alternatively Eqs.~\eqref{eq:LaplaceEquationModified_RHS_RME} and \eqref{eq:monopoledipoleIntensity_FINAL} until convergence is reached, as defined by the following criterion between two successive iterations:
\begin{equation}
    \left\| \frac{ \langle c\boldsymbol{n}\rangle^{k+1}-\langle c\boldsymbol{n}\rangle^{k}}{\langle c\boldsymbol{n}\rangle^{k+1}}\right\|_{\infty} < \epsilon,
    \label{eq:approximateRelativeError}
\end{equation}
where $\langle c\boldsymbol{n}\rangle^{k}$ is the vector collecting the polarities of the $N$ particles at iteration $k$.
For the results presented in this work, we set the tolerance to $\epsilon=10^{-10}$ in our calculations.

\subsubsection{Regularized moments of the concentration distribution}
Finding the dipole intensity, $\boldsymbol{q}^{D}_n$,  requires computing the polarity $\langle c \boldsymbol{n} \rangle_n$ which is in principle defined \emph{at the particle's surface}. To follow the spirit of FCM, and allow for efficient numerical treatment, this surface projection is replaced by a weighted projection over the entire volume $V_F$:
\begin{equation}
    \langle c \boldsymbol{n} \rangle_n =\frac{1}{4\pi a^2} \int_{S_n} c \boldsymbol{n} \mathrm{d}S \quad\longrightarrow\quad \{ c \boldsymbol{n} \}_n = \int_{V_F} c \boldsymbol{n}_n \Delta^P(\rb_n) \mathrm{d}V,
    \label{eq:polarityFCM_FINAL}
\end{equation}
with $\boldsymbol{n}_n$ now defined as $\boldsymbol{n}_n={\boldsymbol{r}_n}/{r_n}$, and the regular averaging kernel $\Delta^P$ for the polarity as:
\begin{equation}
    \Delta^P(\rb) = \frac{r}{8\pi\sigma_P^4} \ \mathrm{exp}\left(-\frac{r^2}{2 \sigma_P^2}\right).
    \label{eq:averagingEnvelopPolarity_FINAL}
\end{equation}
Beyond its importance for determining the dipole intensity associated to a given particle, we will later show that the polarity of the concentration at particle $n$'s surface is directly related to its self-induced phoretic velocity, Eq.~\eqref{eq:translationalAndRotationalVelocitiesCoupling}, and that, similarly, the self-induced hydrodynamic stresslet signature of the particle is in general associated to the first two moments of the surface concentration.
Similarly to the polarity, the second surface moment, $\langle c(\boldsymbol{n}\boldsymbol{n}-\mathbf{I}/3) \rangle_n$ will be replaced in our implementation by a weighted volume projection $\{ c(\boldsymbol{n}\boldsymbol{n}-\mathbf{I}/3) \}_n$:
\begin{equation}
    \langle c(\boldsymbol{n}\boldsymbol{n}-\mathbf{I}/3) \rangle_n=\frac{1}{4\pi a^2}\int_{S_n}c\left(\nb\nb-\frac{\mathbf{I}}{3}\right)\mathrm{d}S \rightarrow \{ c(\boldsymbol{n}\boldsymbol{n}-\mathbf{I}/3) \}_n = \int_{V_F} c \left(\boldsymbol{n}_n\boldsymbol{n}_n-\frac{\mathbf{I}}{3}\right) \Delta^S(\rb_n) \mathrm{d}V,
    \label{eq:secondMomentOfConcentrationFCM_FINAL}
\end{equation}
where the projection kernel for the second moment of concentration, $\Delta^S$, is defined as:
\begin{equation}
    \Delta^S(\rb)=
    \frac{r^2}{3(2\pi)^{\frac{3}{2}} \sigma_S^5} \ \mathrm{exp}\left(-\frac{r^2}{2 \sigma_S^2}\right).
    \label{eq:averagingEnvelopeSecondMoment_FINAL}
\end{equation}

\begin{figure}
    \begin{center}
    \includegraphics[width=0.5\textwidth]{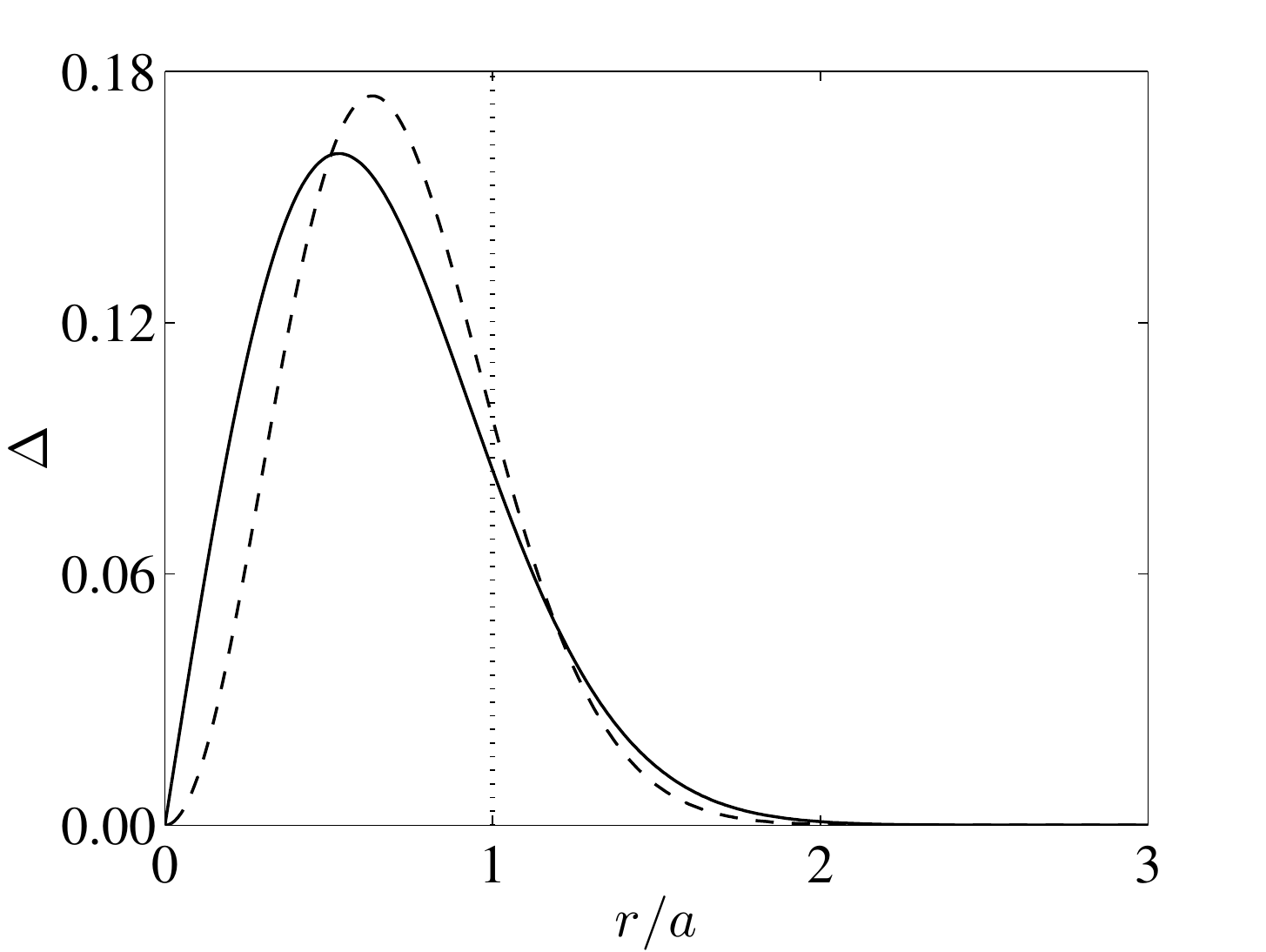}
    \end{center}
    \caption{Averaging envelopes for the first and second moments of concentration, $\Delta^P$ (solid, Eq.~\eqref{eq:averagingEnvelopPolarity_FINAL}) and $\Delta^S$ (dashed, Eq.~\eqref{eq:averagingEnvelopeSecondMoment_FINAL}) respectively. The numerical values for $\sigma_P$ and $\sigma_S$ are set from Eqs.~\eqref{eq:sigmaPsigmaDvalues_FINAL} and \eqref{eq:sigmaSsigmaQvalues_FINAL}.}
    \label{fig:averagingEnvelopes}
\end{figure}
The envelopes $\sigma_P$ and $\sigma_S$ are free parameters in the method that need to be calibrated.
In our reactive FCM formulation, we use modified forms of the Gaussian operator $\Delta$ as projection operators, Eqs.~\eqref{eq:averagingEnvelopPolarity_FINAL} and \eqref{eq:averagingEnvelopeSecondMoment_FINAL}, in order to ensure a fast numerical convergence of the integration for the first and second moments calculation, Eqs.~\eqref{eq:polarityFCM_FINAL} and \eqref{eq:secondMomentOfConcentrationFCM_FINAL} respectively. The integrals over the entire volume $V_F$ of these averaging functions is still equal to one, and their weight is shifted from the particle centre and toward the particle surface (figure~\ref{fig:averagingEnvelopes}), which is both numerically more accurate and more intuitive physically  as these operators are used to obtain the properties of the particle on their surface.

\subsubsection{Calibrating the spreading/averaging envelopes.}
\label{sec:calibrating_env}
Our method relies on four numerical parameters ($\sigma_M$, $\sigma_D$, $\sigma_P$, $\sigma_S$) that we choose to calibrate so as to ensure that several key results in reference configurations are obtained exactly. In particular, to properly account for the phoretic drift induced by the other particles, we ensure that the polarity $\langle c\nb\rangle$ of an isolated  particle placed in an externally-imposed uniform gradient of concentration can be exactly recovered using the regular representation and averaging operators. A similar approach is then followed for the particle's second moment of concentration $\langle c(\nb\nb-\mathbf{I}/3)\rangle$ in a quadratic externally-imposed field.\\

\textbf{Isolated passive particle in an external linear field} -- 
We first consider a single particle placed at the origin in an externally-imposed linear concentration field so that for $r\gg a$, $c\approx c_E$ with
\begin{equation}
    c_E = \boldsymbol{L}_E \cdot \boldsymbol{r},
    \label{eq:externalLinearField}
\end{equation}
where $\boldsymbol{L}_E$ is the externally-imposed uniform gradient. For a passive particle (i.e. $\alpha=0$), satisfying the boundary condition, Eq.~\eqref{eq:boundaryConditionDiffusionSphereN}, at the surface of the particle imposes that the exact concentration distribution around the particle is $c=c_E+c_I^o$, with ${c_I^o(\rb)=a^3\boldsymbol{L}_E\cdot\rb/(2r^3)}$ a singular dipole induced field. The polarity of the external and induced parts, $c_E$ and $c_I$, can be obtained analytically as:
\begin{equation}
    \langle c_E \boldsymbol{n} \rangle = \frac{a}{3}\boldsymbol{L}_E,\qquad
    \langle c_I^o \boldsymbol{n} \rangle = \frac{a}{6}\boldsymbol{L}_E.
    \label{eq:singularFirstMomentsLinearField}
\end{equation}
Following the framework presented above, the regularized solution can be written $c=c_E+c_I^r$ with $c_i^r$ a regularized dipole, and the corresponding regularized-volume moments based on Eq.~\eqref{eq:polarityFCM_FINAL} are obtained using Eq.~\eqref{eq:DipoleregularizedSolution}, as
\begin{equation}
    \{ c_E \boldsymbol{n} \} = \sqrt{\frac{\pi}{8}} \sigma_P \boldsymbol{L}_E,\qquad
    \{ c_I^r \boldsymbol{n} \} = \frac{a^3\sigma_P}{12(\sigma_D^2+\sigma_P^2)^{\frac{3}{2}}} \  \boldsymbol{L}_E.
    \label{eq:regularizedFirstMomentsLinearField}
\end{equation}
Identification of the regularized result \eqref{eq:regularizedFirstMomentsLinearField} to the true solution \eqref{eq:singularFirstMomentsLinearField}, determines $\sigma_P$ and $\sigma_D$ uniquely as:
\begin{equation}
    \frac{\sigma_P}{a}=\frac{1}{3}\sqrt{\frac{8}{\pi}} \approx 0.5319, \qquad \frac{\sigma_D}{a} = \sqrt{\big(\frac{\sigma_P}{2a}\big)^{2/3} - \big(\frac{\sigma_P}{a}\big)^{2}} \approx 0.3614.
    \label{eq:sigmaPsigmaDvalues_FINAL}
\end{equation}
\\

\textbf{Isolated passive particle in an external quadratic field} --
Similarly, in an external quadratic field $c_E$ of the form:
\begin{equation}
    c_E(\rb)= \boldsymbol{r} \cdot \mathbf{Q}_E \cdot \boldsymbol{r},
    \label{eq:externalQuadraticField}
\end{equation}
with $\mathbf{Q}_E$ a  second-order symmetric and traceless tensor, the concentration distribution around a passive particle ($\alpha=0$) takes the form $c=c_E+c_I^o$ with $c_I^o(\rb)$ an induced singular quadrupole.  The exact and regularized second moments of the external field $c_E$ at the particle surface {are} equal to 
\begin{equation}
    \langle c_E (\boldsymbol{n}\boldsymbol{n}-\mathbf{I}/3) \rangle = \frac{2a^2}{15}\mathbf{Q}_E,	\qquad \{c_E (\boldsymbol{n}\boldsymbol{n}-\mathbf{I}/3) \} = \frac{2\sigma_S^2}{3}\mathbf{Q}_E.
    \label{eq:singularSecondMomentsQuadraticField}
\end{equation}

Identifying both results determines the size of the averaging envelope for the second moment uniquely, as
\begin{equation}
    \frac{\sigma_S}{a}=\sqrt{\frac{1}{5}} \approx 0.4472.
    \label{eq:sigmaSsigmaQvalues_FINAL}
\end{equation}
Note that we do not enforce here a constraint on the representation of the second moment of the induced field $c_I$, since the particles' representation do not include a regularized quadrupole in our method.\\

The value $\sigma_M$ remains as a free parameter at this point and cannot be calibrated with a similar approach. In the following, in order to minimize the number of distinct numerical parameters and to minimize the departure of the regularized solution from its singular counterpart, we set its value equal to the smallest envelope size, namely $\sigma_M=\sigma_D$. These specific values of the parameters were used in figures \ref{fig:greensFunctionsFCMandSME} and \ref{fig:averagingEnvelopes}.

\subsection{Hydrodynamic FCM}\label{sec:hydroFCM}
To compute the hydrodynamic interactions between phoretic particles, we rely on the Force Coupling Method (FCM).
This section briefly describes the existing FCM framework developed for the simulation of passive and active suspensions in Stokes flow.

\subsubsection{FCM for passive suspensions}
With hydrodynamic FCM, the effect of the particles on the fluid is accounted for through a forcing term $\boldsymbol{f}$ applied to the dimensionless Stokes equations 
\begin{equation}
    \nabla p - \nabla^2 \boldsymbol{u}= \boldsymbol{f}(\boldsymbol{r},t) \quad \quad \mathrm{in} \ V_F.
    \label{eq:StokesEquation1FCM}
\end{equation}
As for reactive FCM, this forcing arises from a truncated regularized multipolar expansion up to the dipole level
\begin{equation}
    \boldsymbol{f}(\boldsymbol{r},t) = \sum_{n=1}^N \Big[ \boldsymbol{F}_n \Delta (r_n) +  \mathbf{D}_n \cdot \nabla \Delta^*(r_n) \Big],
    \label{eq:RHSStokesEquationFCM}
\end{equation}
where the spreading envelopes are defined by
\begin{equation}
    \Delta(r) = (2\pi\sigma^2)^{-3/2} \mathrm{exp}\left(-\frac{r^2}{2 \sigma^2}\right), \qquad
    \Delta^*(r)  = (2\pi\sigma_*^2)^{-3/2} \mathrm{exp}\left(-\frac{r^2}{2 \sigma_*^2}\right).
    \label{eq:envelopeMonopoleDipoleHydrodynamics}
\end{equation}
$\boldsymbol{F}_n$ and $\mathbf{D}_n$ are the force monopole and dipole applied to particle $n$. The force dipole can be split into a symmetric part,  the stresslet $\mathbf{S}$, and an antisymmetric one related to the external torque $\boldsymbol{T}$:
\begin{equation}
    \mathbf{D}_n = \mathbf{S}_n  + \frac{1}{2} \boldsymbol{\epsilon} \cdot \boldsymbol{T}_n,
    \label{eq:hydrodinamicDipoleDecomposition}
\end{equation}
with $\boldsymbol{\epsilon}$ the third-order permutation tensor. The corresponding regularized solution for the fluid velocity $\boldsymbol{u}$ is then obtained as:
\begin{equation}
    \boldsymbol{u} = \boldsymbol{u}(\boldsymbol{r}) = \sum_{n=1}^{N} \big[ \boldsymbol{F}_n \cdot \mathbf{J} (\boldsymbol{r}_n)+ \mathbf{D}_n : \mathbf{R}^*(\boldsymbol{r}_n) \big].
    \label{eq:velocitySolutionFCM}
\end{equation}
For unbounded domains with vanishing perturbations in the far-field (i.e. $\|\boldsymbol{u}\|\rightarrow 0$ when $r \rightarrow \infty$), the regularized Green's function $\mathbf{J}(\boldsymbol{r})$ reads 
\begin{equation}
    \mathbf{J}(\boldsymbol{r}) = \dfrac{1}{8 \pi r} \left(A(r) \mathbf{I} + B(r) \dfrac{\boldsymbol{r}\boldsymbol{r}}{r^2}\right),
    \label{eq:StokesletGreenFunction}
\end{equation}
with
\begin{align}
    A(r)&= \left(1+\frac{\sigma^2}{r^2}\right) \mathrm{erf}\left(\frac{r}{\sigma \sqrt{2}}\right) - \frac{\sigma}{r} \sqrt{\frac{2}{\pi}} \mathrm{exp}\left(-\frac{r^2}{2\sigma^2}\right),
    \label{eq:functionA}\\
    B(r)&=  \left(1-\frac{3\sigma^2}{r^2}\right) \mathrm{erf}\left(\frac{r}{\sigma \sqrt{2}}\right) + \frac{3\sigma}{r}\sqrt{\frac{2}{\pi}}  \mathrm{exp}\left(-\frac{r^2}{2\sigma^2}\right),
    \label{eq:functionB}
\end{align}
and $\mathbf{R}^*=\nabla \mathbf{J}^*$  is the FCM dipole Green's function evaluated with the parameter $\sigma_*$.

The particle's translational and angular velocities, $\boldsymbol{U}_n$ and $\boldsymbol{\Omega}_n$, are obtained from a volume-weighted average of the local fluid velocity and vorticity
\begin{equation}
    \boldsymbol{U}_n = \int_{V_F} \boldsymbol{u} \, \Delta(\rb_n) \mathrm{d}V, \qquad \boldsymbol{\Omega}_n = \frac{1}{2}\int_{V_F} [\nabla \times \boldsymbol{u}] \Delta^*(\rb_n) \mathrm{d}V.
    \label{eq:meanTranslationalVelocityFCM}
\end{equation}
The Gaussian parameters, $\sigma$ and $\sigma^*$ are calibrated  to recover the correct Stokes drag, ${\boldsymbol{F}=6\pi a \mu \boldsymbol{U}}$, and viscous torque, $\boldsymbol{T}=8\pi a^3 \mu \boldsymbol{\Omega}$, of an isolated particle~\citep{MaxeyPatel2001,LomholtMaxey2003}, leading to 
\begin{equation}
    \frac{\sigma}{a} =\frac{1}{\sqrt{\pi}} \approx 0.5641,  \qquad \qquad  \frac{\sigma_*}{a}=\frac{1}{(6\sqrt{\pi})^{1/3}} \approx 0.4547.
    \label{eq:envelopeSizesHydrodynamics}
\end{equation}

The rigidity of the particle is similarly weakly enforced by imposing that the volume-averaged strain rate $\mathbf{E}_n$ over the envelope of particle $n$ vanishes:
\begin{equation}
    \mathbf{E}_n = \frac{1}{2} \int_{V_F} [\nabla \boldsymbol{u} + (\nabla \boldsymbol{u})^\mathrm{T}] \Delta^*(\rb_n)\mathrm{d}V = \boldsymbol{0},
    \label{eq:meanRateOfStrainFCM}
\end{equation}
which determines the stresslet $\mathbf{S}_n$ induced by particle $n$. Note that unlike forces and torques which are typically set by external or inter-particle potentials, the stresslets  result from the constraint on the flow given by Eq.~\eqref{eq:meanRateOfStrainFCM} and, consequently, need to be solved for as part of the general flow problem. The resulting linear system for the unknown stresslet coefficients is solved directly or iteratively, with the conjugate gradients method, depending on the number of particles considered \citep{LomholtMaxey2003,YeoMaxey2010}. In the following, we consider pairs of particles (see Section \ref{sec:Results})  and therefore use direct inversion.

Note that the averaging envelopes used to recover the translational  and rotational  velocities, $\triangle_n$ and $\triangle^*_n$, are exactly the same as the spreading operators in \eqref{eq:RHSStokesEquationFCM}, all of them Gaussian functions. As a result, the spreading and averaging operators are adjoints to one another. Also note that only two envelope lengths are required for the  hydrodynamic problem: $\sigma$ and $\sigma_*$. 
In contrast, the new reactive FCM extension presented in Section~\ref{sec:ReactiveFCM} uses spreading and averaging operators  that are not adjoint.  To recover the first (\ref{eq:polarityFCM_FINAL}) and second (\ref{eq:secondMomentOfConcentrationFCM_FINAL}) moments of concentration we have two non-Gaussian averaging envelopes ($\Delta^P$ and $\Delta^S$), that differ from the  Gaussian spreading envelopes ($\Delta^M$ and $\Delta^D$) in (\ref{eq:LaplaceEquationModified_RHS_RME}). While having adjoint operators is crucial in hydrodynamic FCM to satisfy the fluctuation-dissipation balance, the lack of adjoint properties for the Laplace problem does not raise any issue in the deterministic setting.

\subsubsection{Active hydrodynamic FCM}
In recent years, FCM has been extended to handle suspensions of active particles, such as microswimmers.
In addition to undergoing rigid body motion in the absence of applied forces or torques, active and self-propelled particles are also characterized by the flows they generate. These flows can be incorporated into FCM by adding an appropriate set of regularized multipoles to the Stokes equations. This problem was solved previously for the classical squirmer model~\citep{DelmotteAllCliment2015}, a spherical self-propelled particle that swims using prescribed distortions of its surface. In the most common case where radial distortions are ignored, the squirmer generates a tangential slip velocity on its surface, just like phoretic particles, which can be expanded into spherical harmonics mode~\citep{Blake1971,Pak2014}.  Consistently with the phoretic problem presented above, only the first two modes are included in the following.

The FCM force distribution produced by $N$ microswimmers self-propelling with a surface slip velocity is given by
\begin{equation}
    \boldsymbol{f}(\boldsymbol{r},t) = \sum_{n=1}^N \Big[ \mathbf{S}_n \cdot \nabla \Delta^*(r_n) + \mathbf{S}^a_n \cdot \nabla \Delta(r_n) + \boldsymbol{H}^a_n \nabla^2 \Delta^*(r_n) \Big],
    \label{eq:RHSStokesEquationForActiveParticlesFCM}
\end{equation}
where $\mathbf{S}^a_n$ is the active stresslet and $\boldsymbol{H}^a_n$ is the active potential dipole associated to the swimming disturbances of swimmer $n$. The latter is defined as 
\begin{equation}
    \boldsymbol{H}^a_n = -2 \pi a^3 \boldsymbol{U}^a_n,
    \label{eq:activeParticleDegenerateQuadrupoleB}
\end{equation}
where $\boldsymbol{U}^a_n$ is the swimming velocity arising from the slip velocity on the swimmer surface $\boldsymbol{u}^s$ \eqref{eq:translationalAndRotationalVelocitiesCoupling}. 
Note that the rigidity stresslet $\mathbf{S}_n$ is included in \eqref{eq:RHSStokesEquationForActiveParticlesFCM} to enforce the absence of deformation of the swimmers, Eq.~\eqref{eq:meanRateOfStrainFCM}. 
The resulting velocity field reads
\begin{equation}
    \boldsymbol{u}(\boldsymbol{r},t) = \sum_{n=1}^{N} \left[ \mathbf{S}_n : \mathbf{R}^*(\rb_n) + \mathbf{S}^a_n : \mathbf{R}(\rb_n) + \boldsymbol{H}^a_n \cdot \mathbf{A}^*(\rb_n) \right],
    \label{eq:velocitySolutionPassiveActiveParticlesFCM}
\end{equation}
where $\mathbf{R}$ is the FCM dipole Green's function   evaluated with the parameter $\sigma$ instead of $\sigma_*$. The second order  tensor $\mathbf{A}^*$ is the FCM Green's function for the potential dipole 
\begin{equation}
   \mathbf{A}^*(\boldsymbol{r}) = \frac{1}{4\pi r^3} \left[ \mathbf{I} - \frac{3\boldsymbol{r}\boldsymbol{r}}{r^2} \right] \mathrm{erf}\left(\frac{r}{\sigma_* \sqrt{2}}\right) - \frac{1}{\mu} \left[ \left( \mathbf{I} - \frac{\boldsymbol{r}\boldsymbol{r}}{r^2} \right) + \left( \mathbf{I} - \frac{3\boldsymbol{r}\boldsymbol{r}}{r^2} \right) \left( \frac{\sigma_*}{r}\right)^2 \right] \Delta^*(r).
    \label{eq:AtensorActiveParticlesFCM}
\end{equation}
The particles' velocity, angular velocity and mean strain rate are then computed as
\begin{align}
    \boldsymbol{U}_n &= \boldsymbol{U}^a_n -\boldsymbol{W}_n + \int_{V_F} \boldsymbol{u} \, \Delta(\rb_n) \mathrm{d}V,
    \label{eq:meanTranslationalVelocityFCM_B} \\
    \boldsymbol{\Omega}_n &= \boldsymbol{\Omega}^a_n + \frac{1}{2}\int_{V_F} [\nabla \times \boldsymbol{u}] \Delta^*(\rb_n) \mathrm{d}V,
    \label{eq:meanRotationalVelocityFCM_B}\\
    \mathbf{E}_n &= -\mathbf{K}_n + \frac{1}{2} \int_{V_F} [\nabla \boldsymbol{u} + (\nabla \boldsymbol{u})^\mathrm{T}] \Delta^*(\rb_n) \mathrm{d}V = \boldsymbol{0},
    \label{eq:meanRateOfStrainFCM_B}
\end{align}
where the active swimming velocities $\boldsymbol{U}^a_n$ and rotation rates $\boldsymbol{\Omega}^a_n$ correspond to the intrinsic velocities of particle $n$, if it was alone (i.e. in the absence of external flows or other particles), and $\boldsymbol{W}_n$ and $\mathbf{K}_n$ are defined as 
\begin{align}
\boldsymbol{W}_n &= \int_{V_F} (\boldsymbol{H}^a_n\cdot\mathbf{A}^*(\rb_n)) \Delta(\rb_n) \mathrm{d}V,\\
\mathbf{K}_n &= \frac{1}{2} \int_{V_F} [\mathbf{S}^a_n:\nabla \mathbf{R}(\rb_n) + (\mathbf{S}_n^a:\nabla \mathbf{R}(\rb_n))^\mathrm{T}] \Delta^*(\rb_n) \mathrm{d}V,
\end{align}
and are included to subtract away the spurious self-induced velocities and local rates of strain arising from the integration of the full velocity field $\boldsymbol{u}$, which already includes the contribution of $\boldsymbol{H}^a_n$ and $\mathbf{S}_n^a$~\citep{DelmotteAllCliment2015}.

\subsection{Diffusio-phoretic FCM}\label{sec:coupling}
At this point, we have described our new reactive FCM framework and have reviewed the key aspects of the existing active hydrodynamic FCM. 
These two steps provide respectively the solution (i) for the concentration field and its moments at the surface of each particles in terms of their position and orientation, and (ii) the particles' velocity in terms of their active hydrodynamic characteristics, i.e. their intrinsic velocities and stresslet, $\boldsymbol{U}_n^a$, $\boldsymbol\Omega_n^a$ and $\mathbf{S}_n^a$. To solve for the full diffusio-phoretic problem (i.e. obtain the  velocity of the particle in terms of their position and orientation), these quantities must be determined from the chemical environment of the particles.
The following section details how to obtain these active characteristics from the output of the reactive problem and provides algorithmic details on the  numerical implementation.
This new diffusio-phoretic framework based on the Force Coupling Method is referred to as DFCM hereafter.

\subsubsection{DFCM: coupling Reactive and Hydrodynamic FCM}
The active swimming speed $\boldsymbol{U}^a_n$ involved in the potential dipole $\boldsymbol{H}^a_n$, \eqref{eq:activeParticleDegenerateQuadrupoleB}, is the phoretic response of particle $n$ to the chemical field, if it was hydrodynamically isolated (i.e. neglecting the presence of other particles in solving the swimming problem). It thus includes its self-induced velocity (i.e. the response to the concentration contrasts induced by its own activity) and the drift velocity induced by the activity of the other particles. The swimming problem for a hydrodynamically-isolated particle in unbounded flows can be solved directly using the reciprocal theorem~\citep{StoneSamuel1996}, and using the definition of the phoretic slip flow
\begin{equation}
    \boldsymbol{U}^a_n \ = \ - \langle \boldsymbol{u}^s \rangle_n \ = \ - \langle M \nabla_{\parallel} c \rangle_n.
    \label{eq:photeticparticleVelocityA}
\end{equation}
After substitution of the mobility distribution at the surface of particle $n$, Eq.~\eqref{eq:trueJanus_1_2_mobiAlteExpr}, using a truncated multipolar expansion of the surface concentration on particle $n$ (up to its second-order moment) and integration by parts, the intrinsic swimming velocity is obtained in terms of the first two surface concentration moments  (see Appendix~\ref{app:velocitycoupling} for more details)
\begin{equation}
    \boldsymbol{U}^a_n = -\frac{2\overline{M}_n}{a} \langle c\boldsymbol{n} \rangle_n
    - \frac{15 M^*_n}{8a} \Big[ 2 \langle c(\boldsymbol{n}\boldsymbol{n}-\mathbf{I}/3) \rangle_n \cdot \boldsymbol{p}_n + \big( \langle c(\boldsymbol{n}\boldsymbol{n}-\mathbf{I}/3) \rangle_n : \boldsymbol{p}_n\boldsymbol{p}_n \big) \, \boldsymbol{p}_n \Big].
    \label{eq:photeticparticleVelocityE}
\end{equation}
Similarly, the active stresslet $\mathbf{S}^a_n$, is defined as in Eq.~\eqref{eq:stressletCoupling},
\begin{align}
    \mathbf{S}^a_n = -10\pi a^2 \langle \boldsymbol{n} \boldsymbol{u}^s +\boldsymbol{u}^s \boldsymbol{n} \rangle_n = -10\pi a^2 \langle M ( \boldsymbol{n} \nabla_{||} c + (\nabla_{||} c) \boldsymbol{n} ) \rangle_n,
    \label{eq:photeticparticleStressletA}
\end{align}
  and rewrites in terms of the moments of concentration (see Appendix~\ref{app:velocitycoupling} for more details) 
\begin{equation}
    \mathbf{S}^a_n = -60 \pi a \overline{M}_n \langle c(\boldsymbol{n}\boldsymbol{n}-\mathbf{I}/3) \rangle_n + \frac{15 \pi a M^*_n}{2} \Big[ (\langle c\boldsymbol{n} \rangle_n \cdot \boldsymbol{p}_n) (\mathbf{I}-\boldsymbol{p}_n\boldsymbol{p}_n) - \langle c\boldsymbol{n} \rangle_n \boldsymbol{p}_n - \boldsymbol{p}_n \langle c\boldsymbol{n} \rangle_n \Big].
    \label{eq:photeticparticleStressletE}
\end{equation}
Finally, the active rotation $\boldsymbol{\Omega}^a_n$, Eq.~\eqref{eq:translationalAndRotationalVelocitiesCoupling}, is obtained in terms of the moments of concentration and the mobility contrast (see Appendix~\ref{app:velocitycoupling})
\begin{equation}
    \boldsymbol{\Omega}^a_n = \frac{9M^*_n}{4a^2} \; \boldsymbol{p}_n \times \langle c\boldsymbol{n} \rangle_n.
    \label{eq:driftRotationalVelocityFCMExpression}
\end{equation}
For uniform mobility, the swimming velocity and stresslet are directly related to the first and second of surface concentrations, but non-uniform mobility introduces a coupling of the different concentration moments. Here, the surface concentration is expanded up to its second-order moment only.

  In our regularized approach, the surface concentration moments appearing in the previous equations will conveniently be computed as weighted volume averages over the entire domain $V_F$ as detailed in Eqs.~\eqref{eq:polarityFCM_FINAL} and \eqref{eq:secondMomentOfConcentrationFCM_FINAL}.\\

Computing the second moment of concentration however requires an additional step: as detailed in Section~\ref{sec:calibrating_env}, the second moment of concentration in an external field arises from the second gradient of that external field, and includes both an externally-induced component $\langle c_E (\boldsymbol{n}\boldsymbol{n}-\mathbf{I}/3) \rangle_n$ (i.e. the moment of that externally-imposed field)  and a self-induced component which corresponds to the second moment of the induced field generated by the particle to ensure that the correct flux boundary condition is satisfied at the particles' surface. For a chemically-inert particle ($\alpha=0$), the self-induced contribution is obtained exactly as  ${\langle c_I^o (\boldsymbol{n}\boldsymbol{n}-\mathbf{I}/3) \rangle_n = \frac{2}{3}\langle c_E (\boldsymbol{n}\boldsymbol{n}-\mathbf{I}/3) \rangle}$. 

Our representation of the particles in the chemical problem is however truncated at the dipole level, Eq.~\eqref{eq:cFieldTruncatedMultipoleExpansion}, and as a result, the quadrupolar response of the particle to the external field can not be accounted for directly. To correct for this shortcoming, we first compute  the external second moment produced by the other particles on particle $n$ using \eqref{eq:secondMomentOfConcentrationFCM_FINAL} and \eqref{eq:cFieldTruncatedMultipoleExpansion},   and multiply the  resulting value by $5/3$ to account for  the full second moment induced by the concentration field indirectly. 

Finally, the particles are themselves active and may generate an intrinsic quadrupole. Its effect on the second surface concentration moment can be added explicitly in terms of the second activity moment, so that the total second moment on particle $n$ is finally evaluated as
\begin{equation}
    \langle c (\boldsymbol{n}\boldsymbol{n}-\mathbf{I}/3) \rangle_n = \frac{5}{3}\{ c_E (\boldsymbol{n}\boldsymbol{n}-\mathbf{I}/3)\}_n + \frac{a}{D}\langle \alpha (\boldsymbol{n}\boldsymbol{n}-\mathbf{I}/3) \rangle_n.
    \label{eq:selfSecondMomentOfConcentration}
\end{equation}

In summary, at a given time step, the particles' velocities are obtained from their instantaneous position and orientation as follows. The first two surface concentration moments are first obtained using our new reactive FCM framework by solving the Poisson problem, Eq.~\eqref{eq:LaplaceEquationModified_RHS_RME}. These moments are then used to compute the phoretic intrinsic translation and rotation velocities, Eqs.~\eqref{eq:photeticparticleVelocityE} and \eqref{eq:driftRotationalVelocityFCMExpression}, as well as the active stresslets and potential dipoles, Eqs.~ \eqref{eq:photeticparticleStressletE} and \eqref{eq:activeParticleDegenerateQuadrupoleB}. The Stokes equations forced by the swimming singularities Eq.~\eqref{eq:RHSStokesEquationForActiveParticlesFCM}, and subject to the particle rigidity constraint, Eq.~\eqref{eq:meanRateOfStrainFCM_B}, are finally solved to obtain the total particle velocities, Eqs.~\eqref{eq:meanTranslationalVelocityFCM_B}--\eqref{eq:meanRotationalVelocityFCM_B}.

\subsubsection{Numerical details}

The volume integrals required to compute the  concentration moments and the hydrodynamic quantities are performed with a  Riemann sum on cartesian grids centred at each particle position. 
To ensure a sufficient resolution, the grid size, $\Delta x$, is chosen so that the smallest envelope size $\sigma_D$ satisfies ${\sigma_{D} = 1.5\Delta x = 0.3614a}$, which corresponds to roughly 4 grid points per radius. 
Owing to the fast decay of the envelopes, the integration domain is truncated so that the widest envelope (that with the largest $\sigma$) essentially vanishes on the boundary of the domain, $\Delta(r) <  \gamma = 10^{-16}$, which, given the grid resolution, requires 39 integration points in each direction. 
Doing so, the numerical integrals yield spectral accuracy.
Setting instead $\gamma = \epsilon = 10^{-10}$, where $\epsilon$ is the relative tolerance for the polarity in the iterative procedure, Eq.~\eqref{eq:approximateRelativeError}, reduces that number to 31 integration points along each axis while keeping a spectral convergence.

\section{Results} \label{sec:Results}
\begin{figure}
    \begin{center}
    \raisebox{1.8in}{}\includegraphics[width=0.30\textwidth]{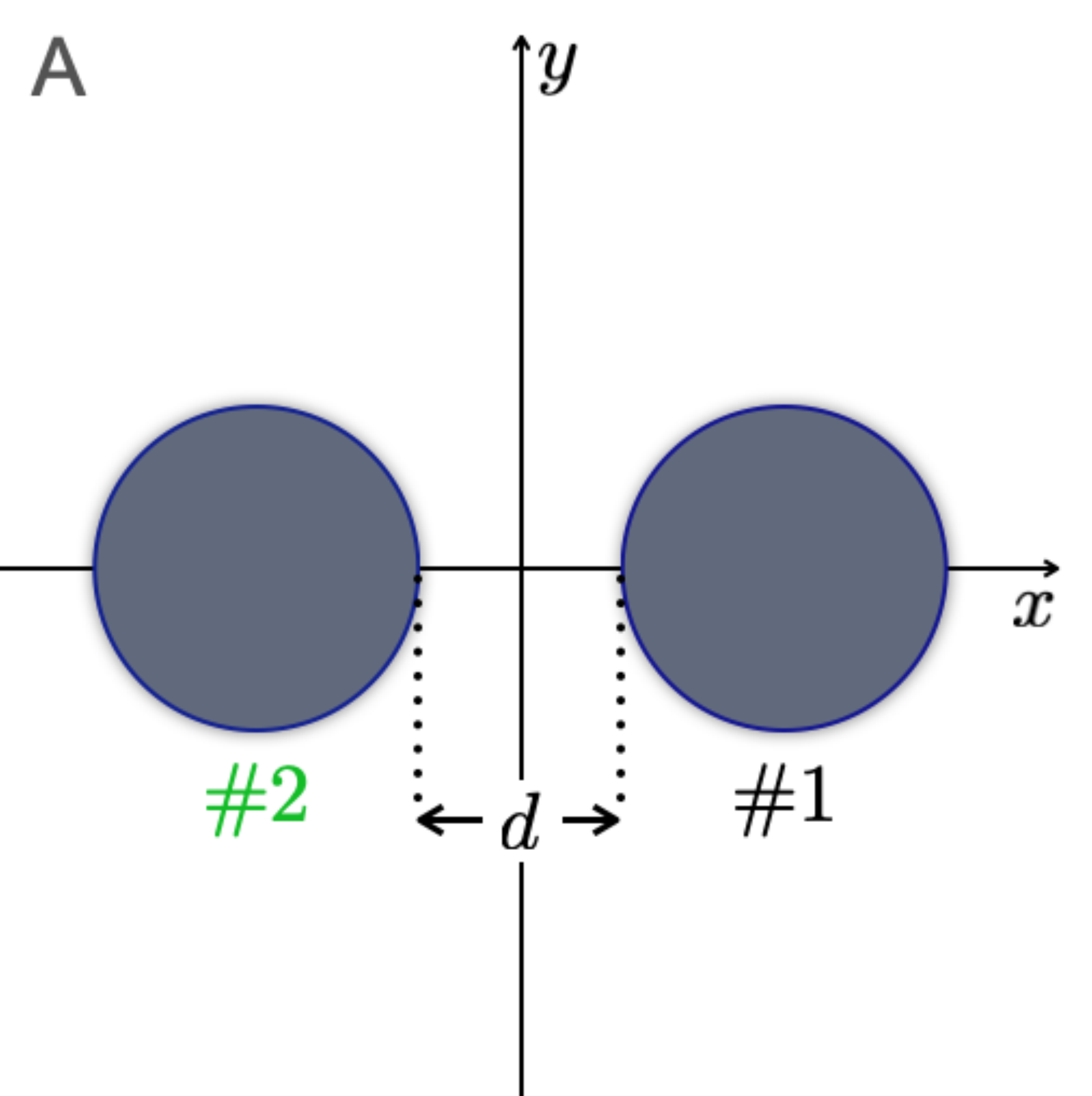}
    \raisebox{1.8in}{}\includegraphics[width=0.30\textwidth]{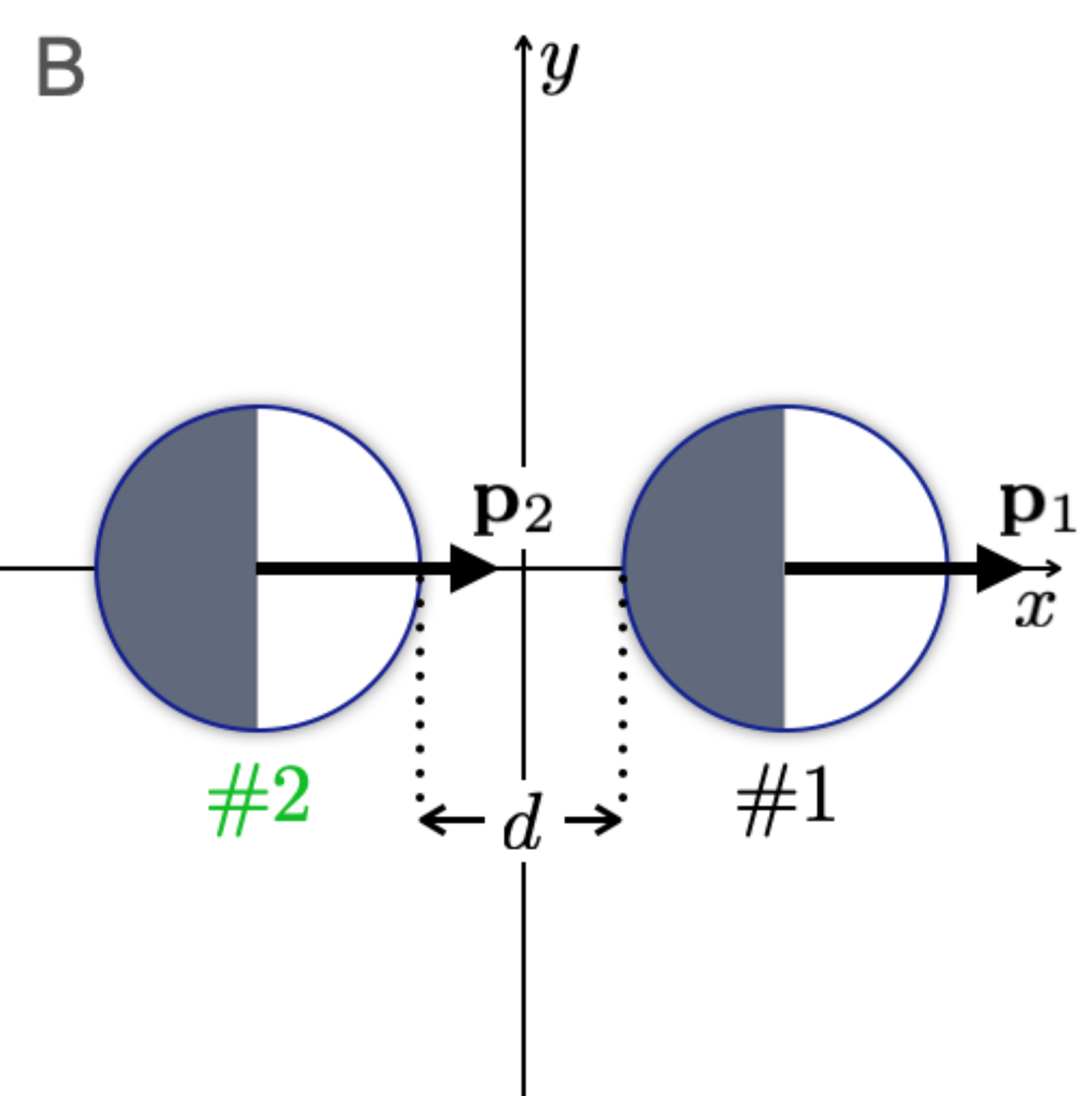}
    \raisebox{1.8in}{}\includegraphics[width=0.30\textwidth]{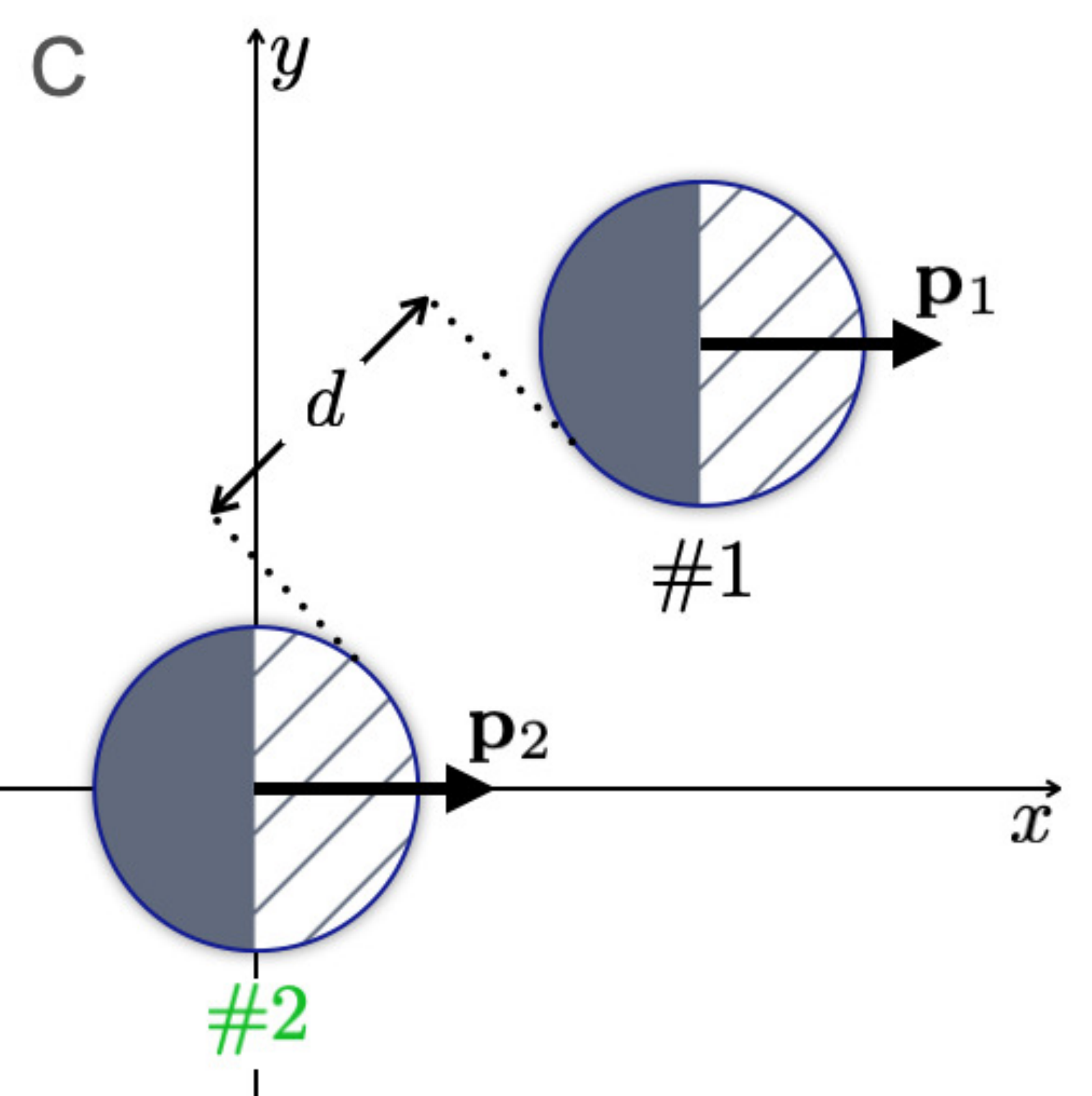}
    \end{center}
    \caption{Validation cases considered: a) Case A: Isotropic particles with uniform mobility, b) Case B:  Hemispheric Janus particles with uniform mobility, c) Case C: Hemispheric Janus particles with non-uniform mobility. In each case, both particles have exactly the same orientation and phoretic properties and their dimensionless separation is noted $d$.}
    \label{fig:ValidationCasesForPublicationCaseACaseBCaseC}
\end{figure}

In this section, we evaluate the accuracy of the present novel DFCM framework  in three different canonical or more generic configurations involving pairs of isotropic and Janus phoretic particles, as shown in figure \ref{fig:ValidationCasesForPublicationCaseACaseBCaseC}. The particles' motion are restricted to a plane within a three-dimensional unbounded domain for the sake of clarity in visualizing the results. 

In this validation process, DFCM is compared with three existing methods providing either a complete or approximate solution of the problem. The simplest one, the Far-Field Approximation model \citep{SotoGolestanian2014,VarmaMichelin2019}, relies on a multipolar expansion of the reactive and hydrodynamic singularities up to the dipole level generated by each particles, but neglects the finite size of the particles (i.e. without reflections on the polarity and rigidity stresslet).
 Our results are also compared to the complete (exact) solution of the problem (i.e. solving the complete hydrodynamic and chemical fields regardless of the particles' distance, accounting for their finite size). For axisymmetric problems, this solution is obtained semi-analytically using the Bi-Spherical Coordinates approach{~\citep{MichelinLauga2015,ReighKapral2015}}, whose   accuracy is only limited by the number of Legendre modes used to represent the solution. For non-axisymmetric configurations, the complete solution is obtained numerically using the regularized Boundary Element Method~\citep{Montenegro-JohnsonMichelinLauga2015}.  These reference solutions are referred to in the following, as FFA, BSC and BEM respectively.
 
\subsection{Isotropic particles - axisymmetric configuration}

\begin{figure}
    \begin{center}   
    \includegraphics[width=0.50\textwidth]{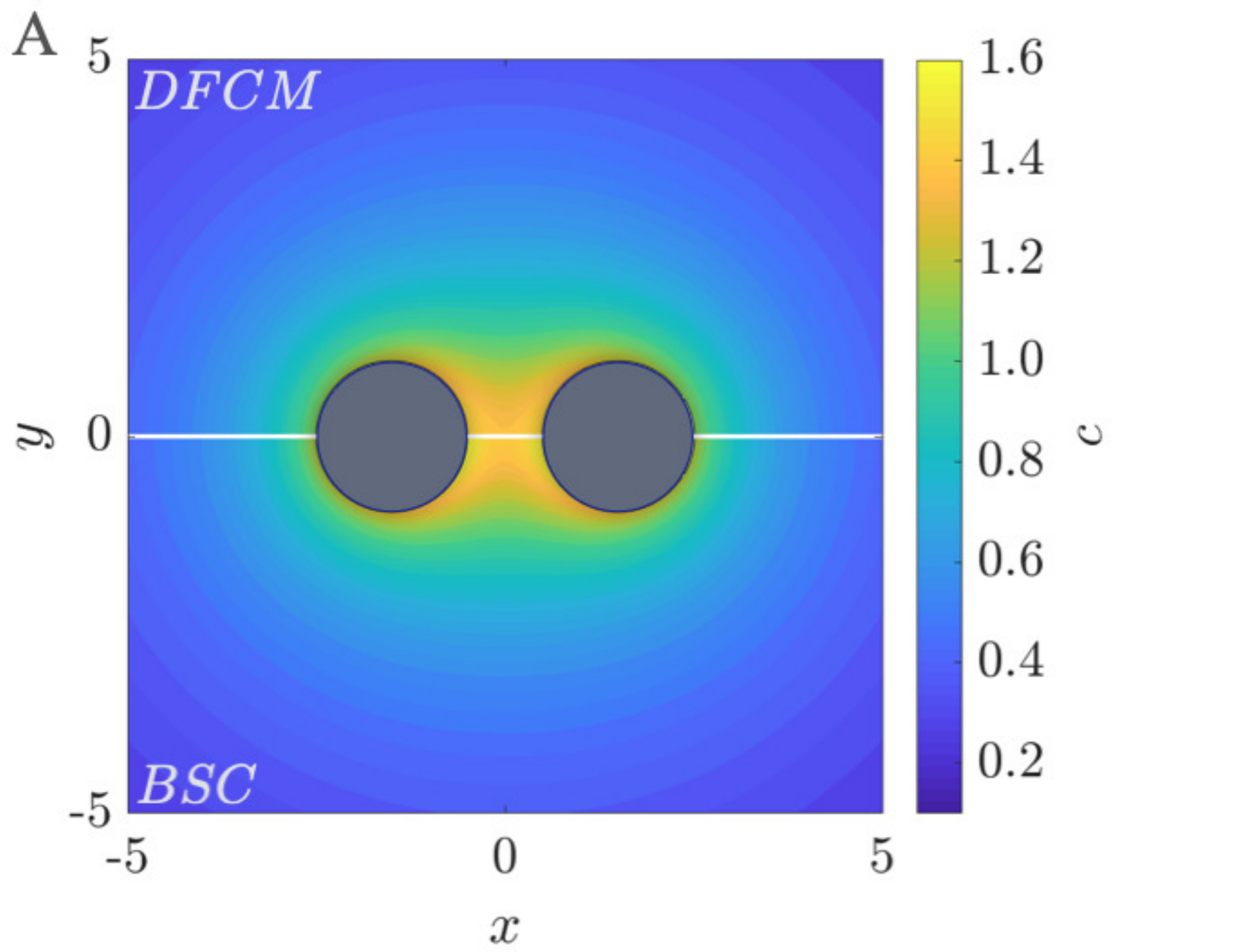}
    \includegraphics[width=0.49\textwidth]{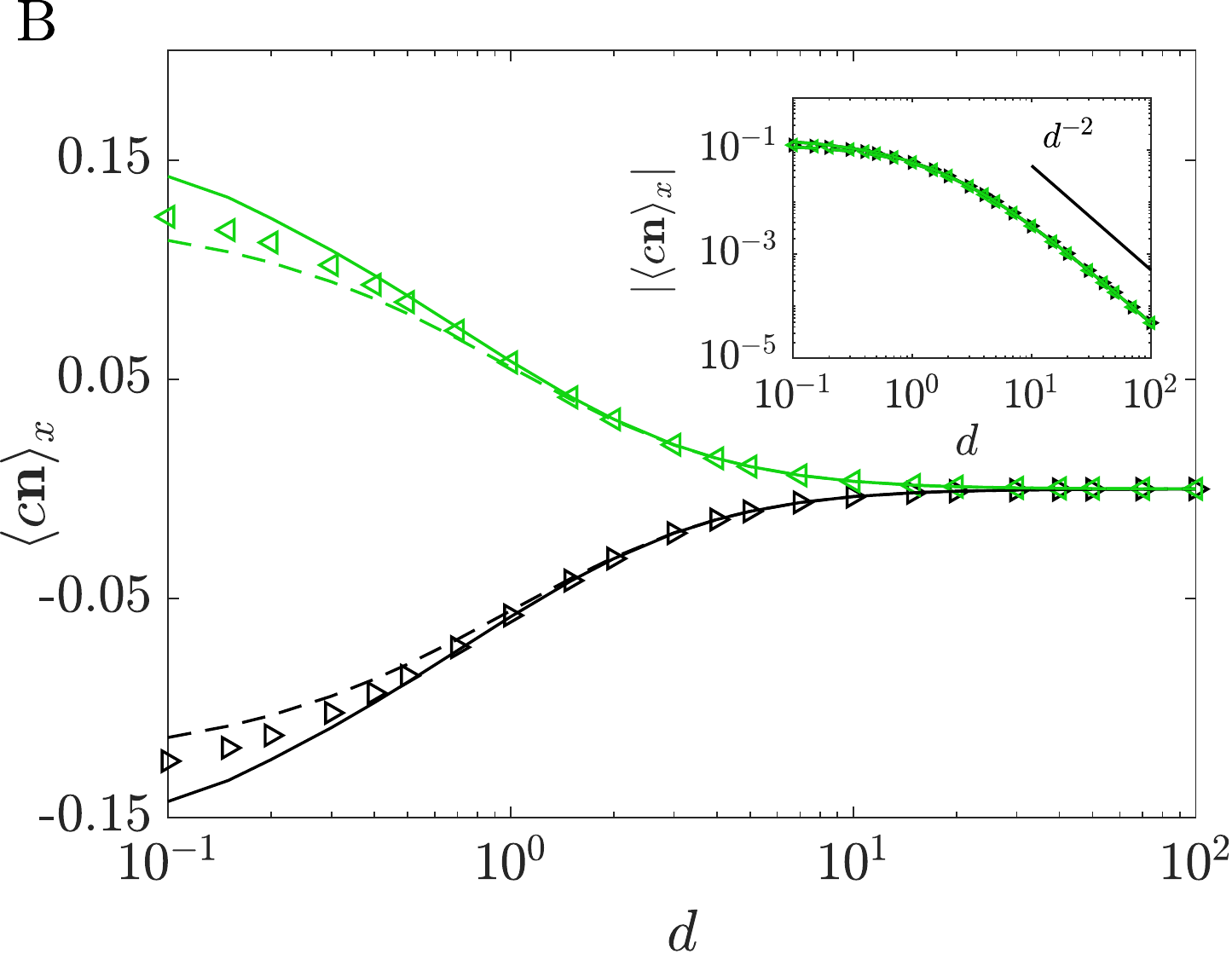}
    \includegraphics[width=0.49\textwidth]{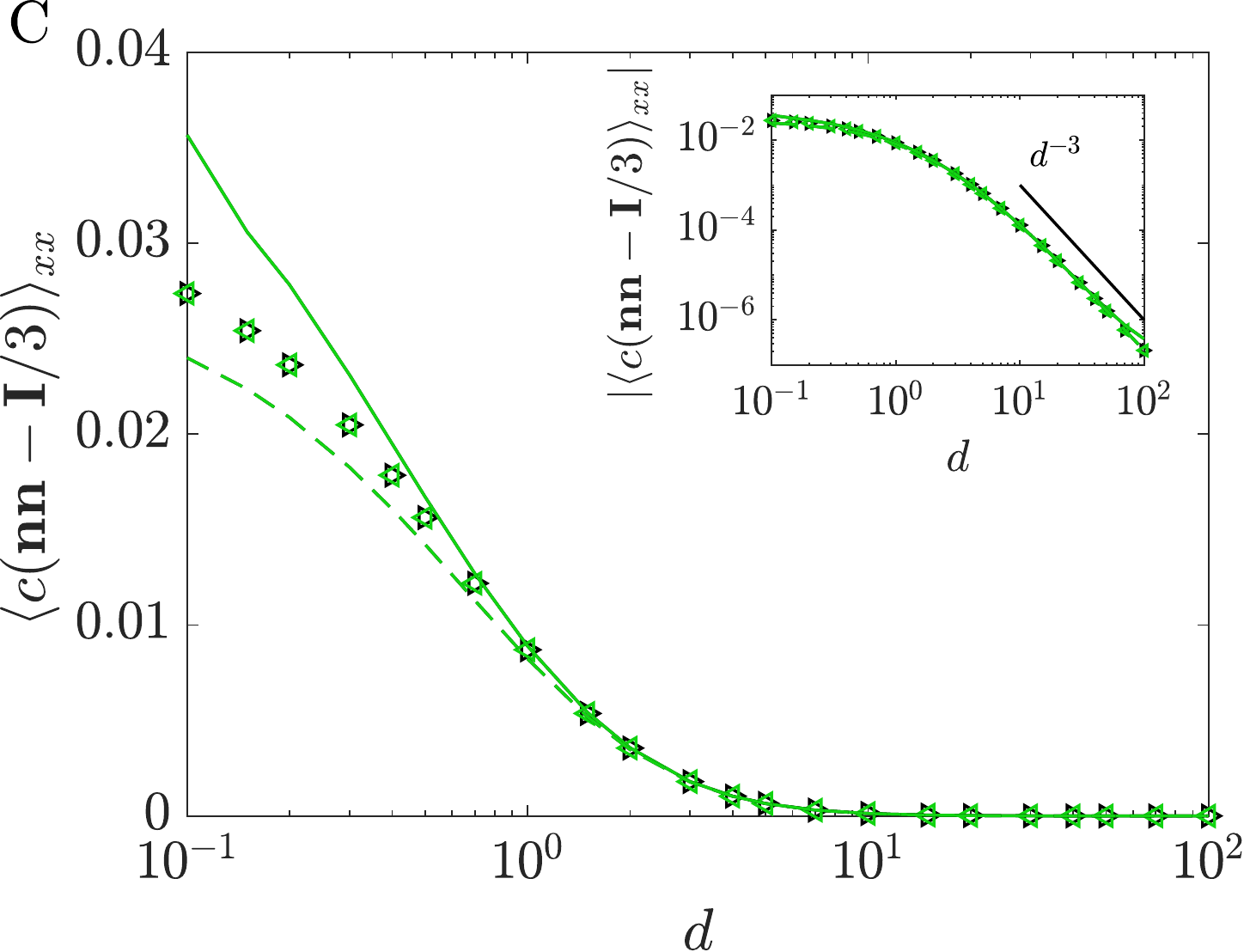}
    \includegraphics[width=0.49\textwidth]{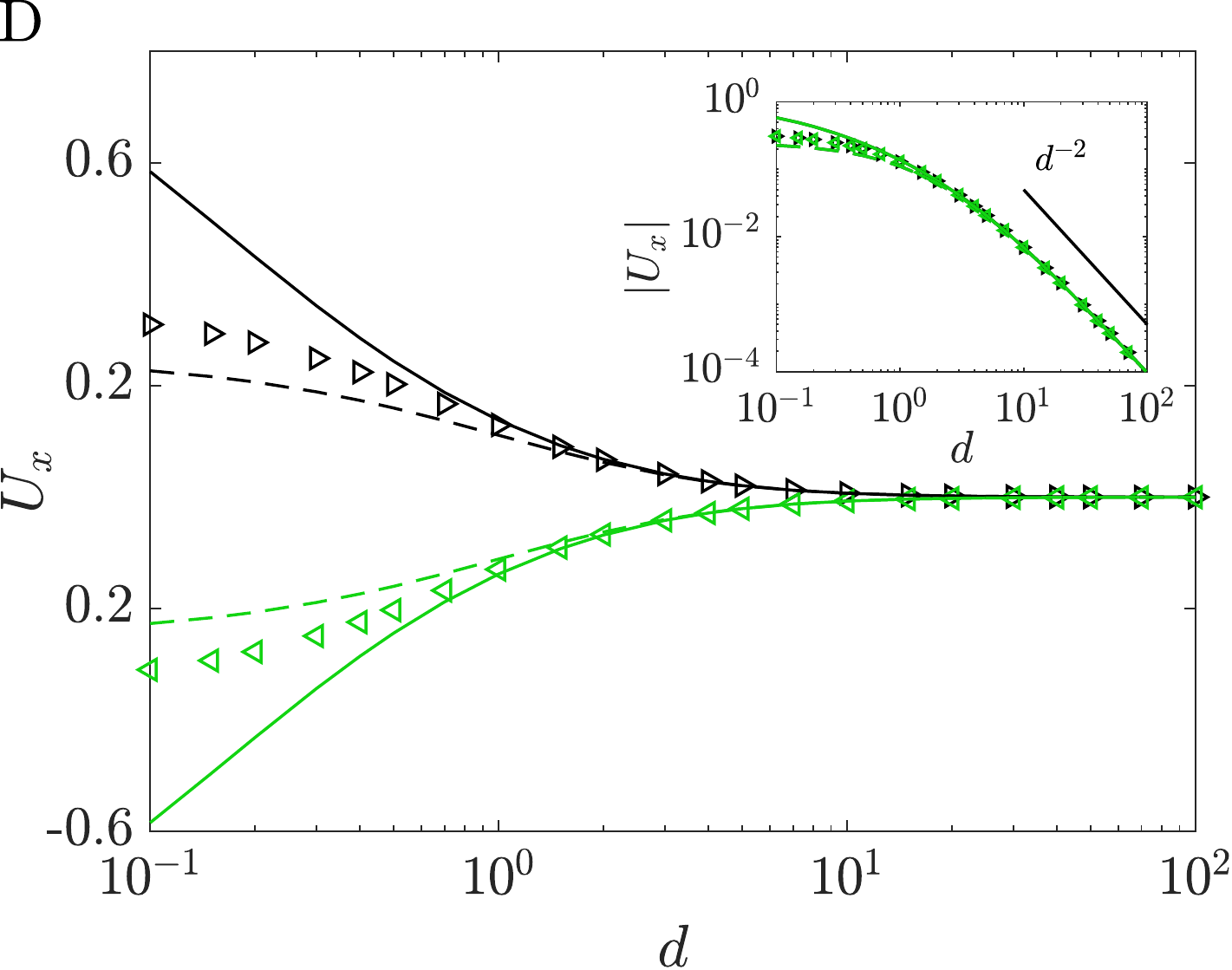}
    \end{center}
    \caption{Case A: a) concentration field for $d=1$ (upper half: DFCM, lower half: BSC),  b) first moment of concentration $\langle c \boldsymbol{n} \rangle_x$,  c) second moment of concentration $\langle c(\boldsymbol{n}\boldsymbol{n}-\mathbf{I}/3) \rangle_{xx}$, d) velocity $U_x$. The black lines (and markers) correspond to particle 1 and the light green ones to particle 2. The triangle markers correspond to DFCM, the solid lines correspond to BSC, while the dashed lines to FFA. The inset shows the absolute values in logarithmic scale and the corresponding decay. The surface averages $\langle ... \rangle$ {were} used for BSC and FFA, while the volume average $\{...\}$ for DFCM. All the omitted components of $\langle c \boldsymbol{n} \rangle$, $\langle c(\boldsymbol{n}\boldsymbol{n}-\mathbf{I}/3) \rangle$, and $\boldsymbol{U}$  are zero.}
    \label{fig:caseAResults}
\end{figure}

The first configuration, Case A (figure~\ref{fig:ValidationCasesForPublicationCaseACaseBCaseC}a), consists of two identical isotropic particles with uniform activity and mobility (${\alpha^F_n=\alpha^B_n=1}$, ${M^F_n=M^B_n=1}$) separated by a distance $d$ along the $x$-axis {\citep{VarmaMontenegro-JohnsonMichelin2018,NasouriGolestanian2020a}}.
Phoretic particles require an asymmetry in their surface concentration field to self-propel \citep{GolestanianLiverpoolAdjari2007}, so that an isolated isotropic particle can not swim. In the configuration considered here however, the concentration gradient produced by a second isotropic particle introduces the required asymmetry to generate motion along the $x$-axis. 

Figure~\ref{fig:caseAResults}(a) shows the concentration field induced by two isotropic particles for $d=1$. The DFCM solution (upper panel) is in good agreement with BSC (lower panel), except near the particles' boundaries in the gap, where the low-order multipolar expansion of DFCM and inaccurate resolution of the particle's surface underestimates the concentration field. The increase in  concentration between the particles is a direct result of the confinement between their active surfaces. It produces a surface concentration gradient and phoretic slip flow on each particle's boundary that pumps the fluid toward this high concentration zone and thus drives the particles away from each other (figure~\ref{fig:caseAResults}d). This effect is magnified as $d$ is reduced, leading to higher particle velocities and  higher moments of concentration for shorter distances. 

The evolution with interparticle distance of the particles' polarity, a measure of the net concentration gradient over their surface,  is shown on figure \ref{fig:caseAResults}(b) as obtained with the DFCM, BSC and FFA approaches. While both FFA and DFCM are in good agreement with the exact solution (BSC) even for relatively small distances, the DFCM approach provides a noticeable improvement over the cruder representation of FFA in the near field  ($d<1$), where the iterative corrections for the mutually-induced polarity  \eqref{eq:approximateRelativeError} contribute significantly.
The expected decay of the polarity as $ 1/d^{2}$ is recovered (figure~\ref{fig:caseAResults}b, inset) in all three cases as the dominant contribution to the polarity is proportional to the gradient of the leading order  monopolar concentration field. Similar results are  obtained  for the second moment of concentration (figure~\ref{fig:caseAResults}c), with an expected $1/d^{3}$-decay proportional to the second gradient of the leading order of the concentration field. We note that isotropic particles do not drive any flow when isolated (and therefore do not have any hydrodynamic signature), but acquire a net stresslet as a result of their chemical interactions, behaving as pusher swimmers.

The resulting translational velocities are shown in figure~\ref{fig:caseAResults}(d): again, 
DFCM performs better than FFA in the range $d<2$ since it additionally considers the hydrodynamic interactions of the particles (e.g. the effect of the rigidity constraint through the rigidity stresslet, see Eq.~\eqref{eq:RHSStokesEquationForActiveParticlesFCM}) in addition to the active flows, while FFA does not.  Such discrepancy arises from the accumulated errors in the successive truncated multipolar expansions: using the BSC solution as a reference, we can determine that for near-field interactions of the two particles around  $25\%-30\%$ of the DFCM error comes from the Reactive FCM approximation \eqref{eq:LaplaceEquationModified_RHS_RME}, while the other $70\%-75\%$ comes from the Hydrodynamical FCM approximation \eqref{eq:RHSStokesEquationForActiveParticlesFCM}.
As expected, in the far-field limit, the velocity decays as $1/d^{2}$ since it is proportional to the polarity to leading order and this dominant contribution does not involve any hydrodynamic interactions: these would correspond at leading order to the contribution of the stresslet generated by the presence of the other particles and decay as $1/d^{5}$~\citep{VarmaMichelin2019}.

\subsection{Janus particles - axisymmetric configuration}
Our second configuration of interest, Case B (figure~\ref{fig:ValidationCasesForPublicationCaseACaseBCaseC}b), focuses on Janus particles, which are currently the most commonly-used configuration for self-propelled phoretic particle in both experiments and theoretical models. Their motion stems from the self-induced concentration gradients produced by the difference in activity between their two hemispheres.
Here we consider two identical Janus particles with uniform mobility ($M^F_n=M^B_n=1$), a passive front cap ($\alpha^F_n=0$) and an active back cap ($\alpha^B_n=1$), leading to a self-propulsion velocity of $\boldsymbol{U}^{\infty}=\frac{1}{4}\boldsymbol{e}_x$~\citep{GolestanianLiverpoolAdjari2007}.  We further focus here on an axisymmetric setting where the particles' orientation coincides with the line connecting their centers, for which an exact semi-analytic solution of the complete hydrochemical problem is available using bispherical coordinates (BSC) as exploited in several recent studies~\citep{VarmaMichelin2019,NasouriGolestanian2020b}. Furthermore, both particles point in the same direction so that, when far enough apart, they swim at the same velocity in the same direction.

\begin{figure}
    \begin{center}
    \includegraphics[width=0.50\textwidth]{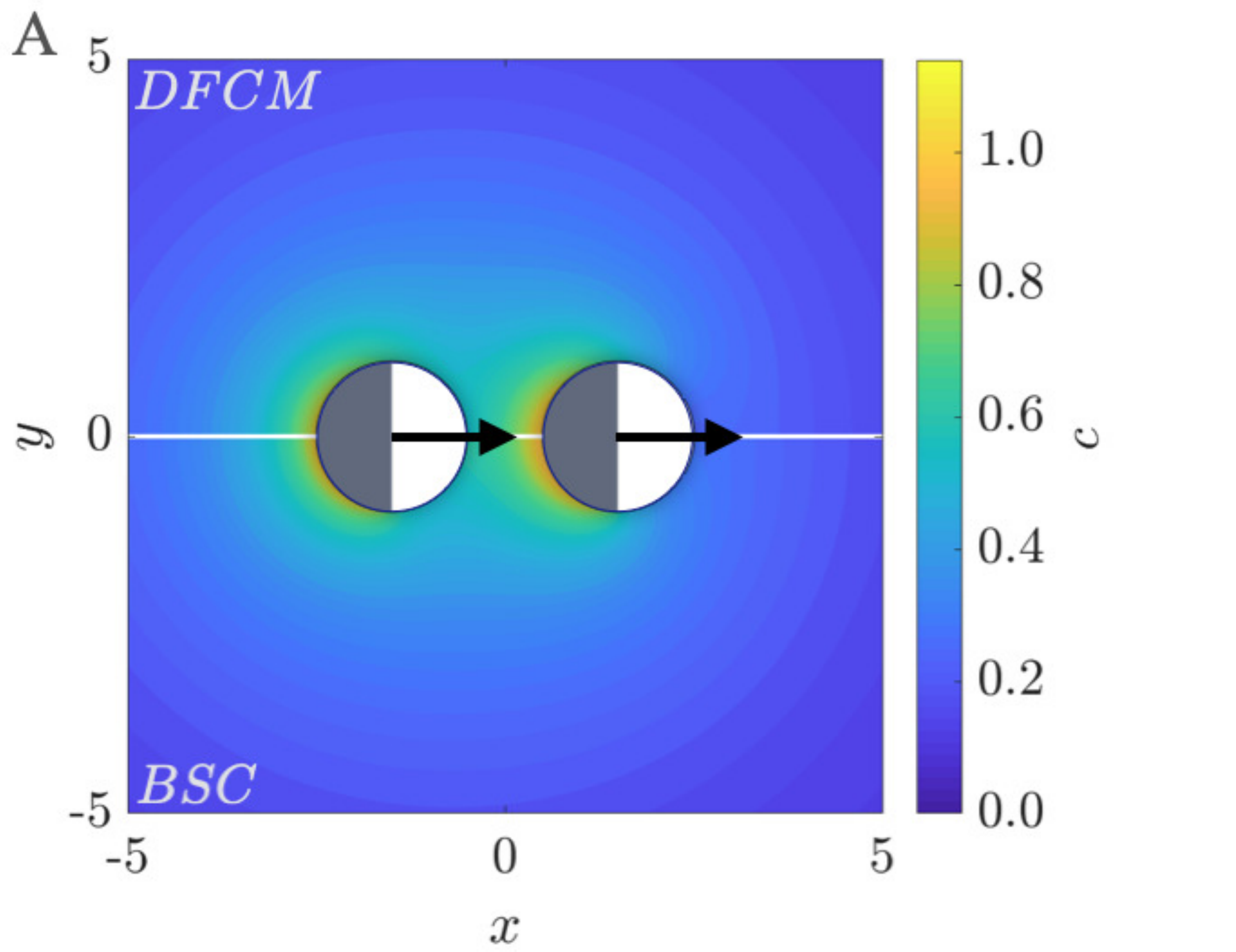}
    \includegraphics[width=0.49\textwidth]{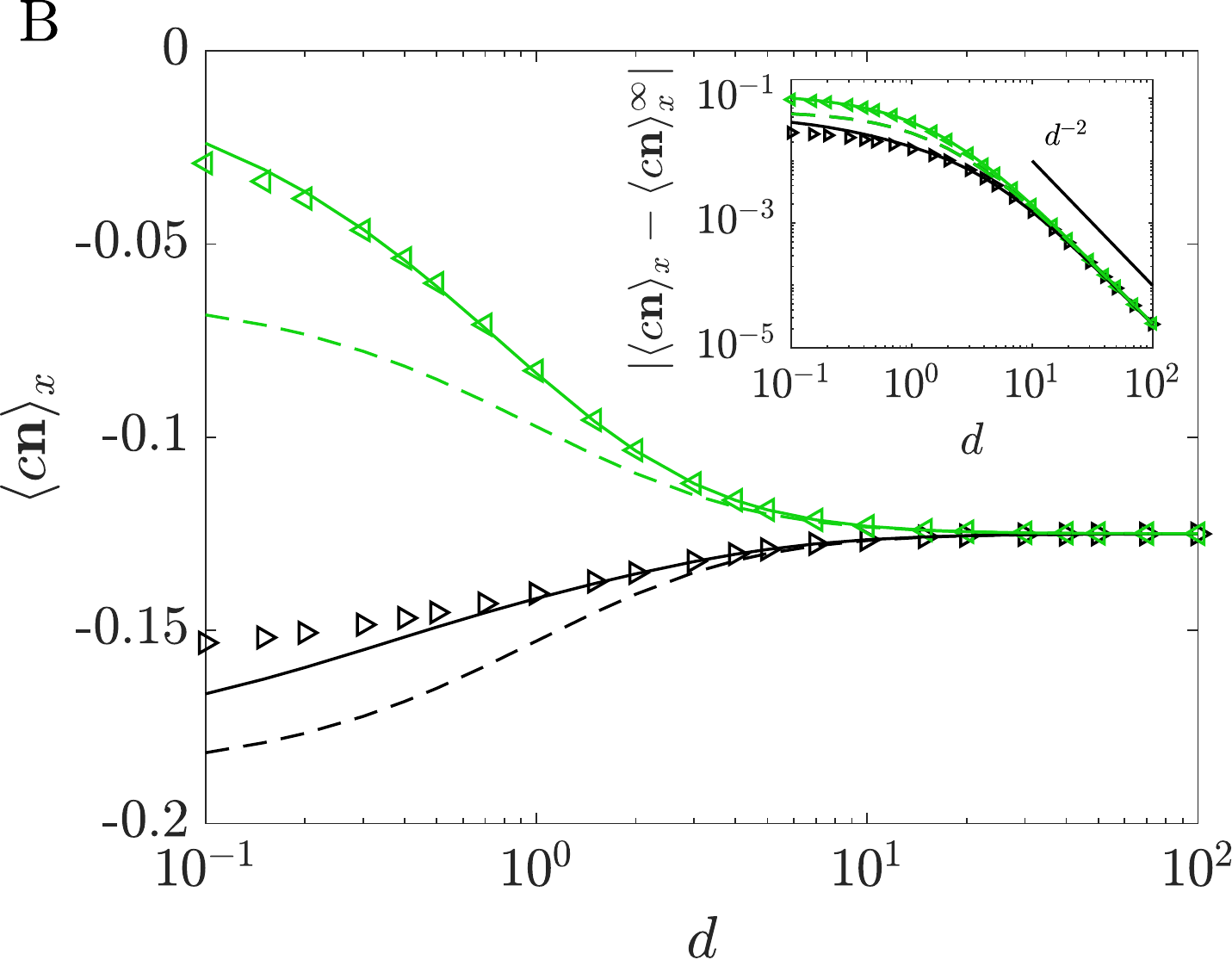}
    \includegraphics[width=0.49\textwidth]{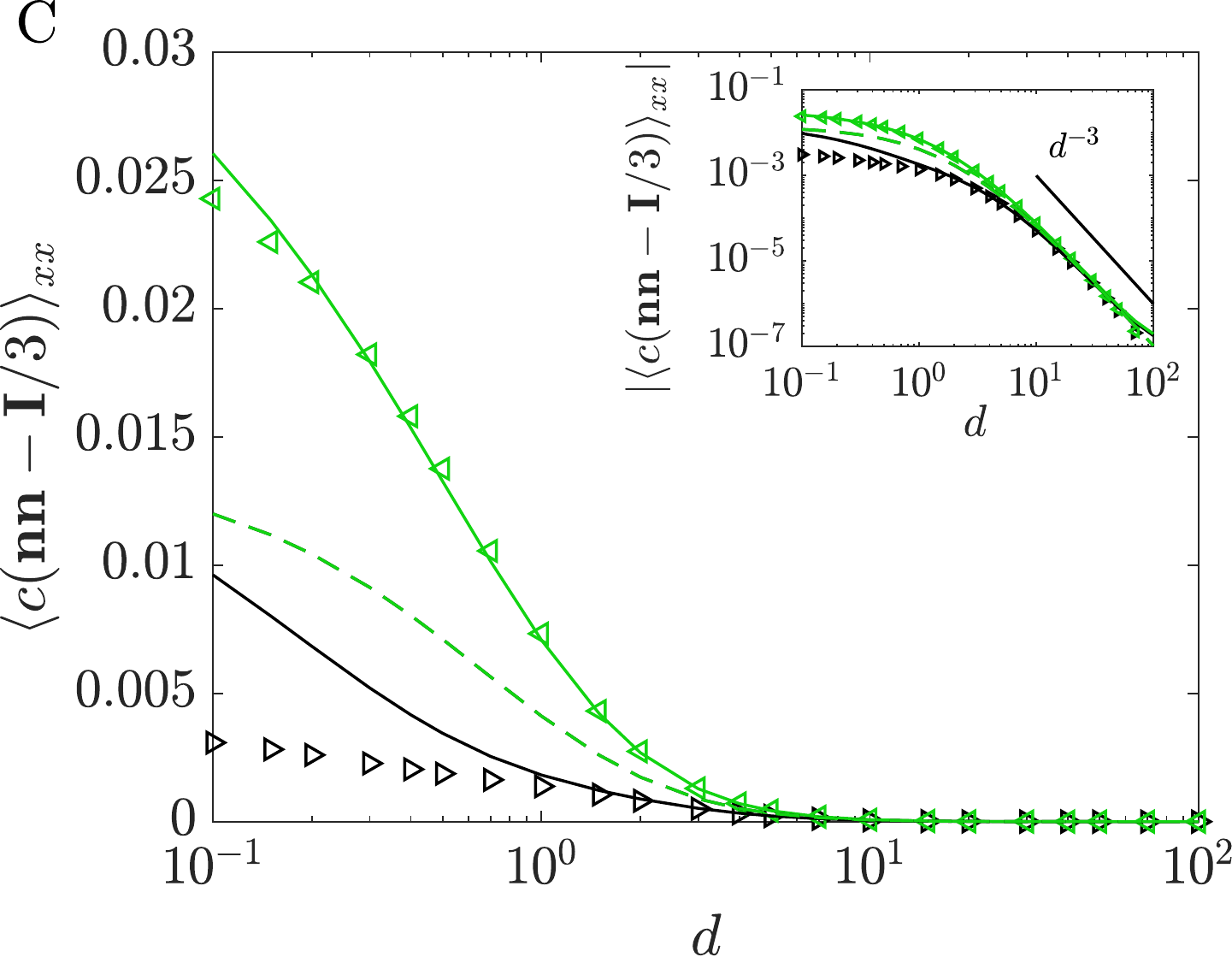}
    \includegraphics[width=0.49\textwidth]{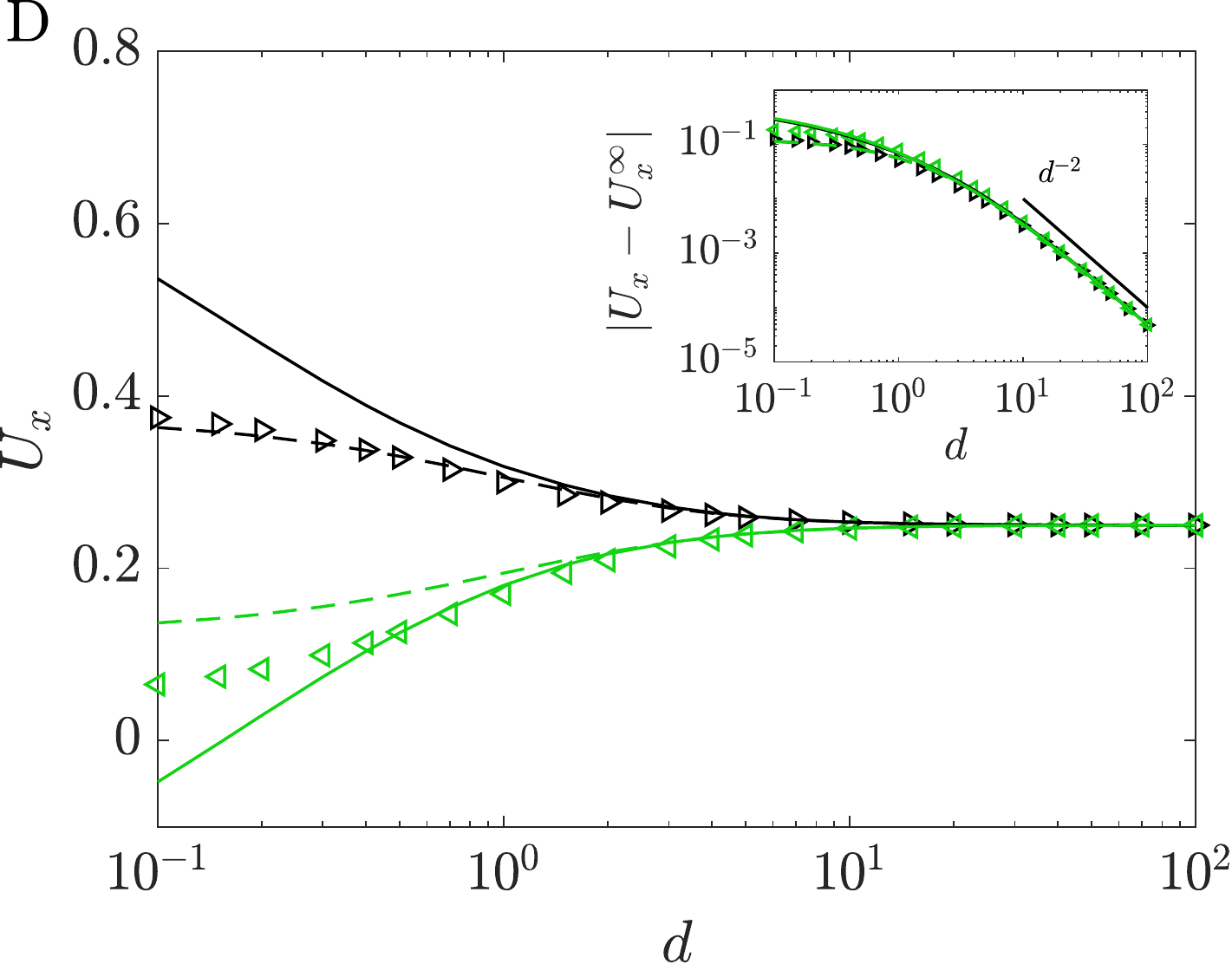}    
    \end{center}
    \caption{Case B: a) concentration field for $d=1$ (upper half: DFCM, lower half: BSC),  b) first moment of concentration $\langle c \boldsymbol{n} \rangle_x$,  c) second moment of concentration $\langle c(\boldsymbol{n}\boldsymbol{n}-\mathbf{I}/3) \rangle_{xx}$, d) velocity $U_x$. The black lines (and markers) correspond to particle 1 and the light green ones to particle 2. The triangle markers correspond to DFCM, the solid lines correspond to BSC, while the dashed lines to FFA. The inset shows the absolute values in logarithmic scale and the corresponding decay. The surface averages $\langle ... \rangle$ where used for BSC and FFA, while the volume average $\{...\}$ for DFCM. All the omitted components of $\langle c \boldsymbol{n} \rangle$, $\langle c(\boldsymbol{n}\boldsymbol{n}-\mathbf{I}/3) \rangle$, and $\boldsymbol{U}$  are zero.}
    \label{fig:caseBResults}
\end{figure}

Figure \ref{fig:caseBResults}(a) shows the concentration field for $d=1$: again, DFCM closely matches the BSC predictions. Here, both particles pump fluid from their front to their active back cap where an excess solute concentration is produced, and therefore move along the $+\boldsymbol{e}_x$ direction. As the interparticle distance  shortens, the concentration increases in the gap, leading to enhanced (resp.\ decreased) surface gradients on the leading (resp. trailing) particle. 

This physical intuition is confirmed by the evolution of the concentration polarity with the interparticle distance (figure \ref{fig:caseBResults}b). The polarity   matches that of an isolated particle $\langle c \boldsymbol{n} \rangle^{\infty}=-\frac{1}{8}\boldsymbol{e}_x$ for large distances $d\gg 1$, and is increased in magnitude for particle 1 (leader) while its magnitude decreases for particle 2 (follower) as $d$ is reduced. The DFCM solution remains in close agreement with BSC for all distances (even down to a tenth of a radius), in particular capturing the asymmetric effect of the interaction on the two particles. In contrast, FFA predicts a symmetric progression of the polarity, leading to large discrepancies for $d<3$.
A similar behaviour is observed for the second moment (figure~\ref{fig:caseBResults}c), except for particle 1 which is underestimated by DFCM in the near field ($d<1$). We note that although isolated Janus particles with uniform mobility behave as neutral {swimmers} (exerting no force dipole or active stresslet on the fluid), their interaction leads to both of them acting as effective pushers on the fluid (negative stresslet, see Eq.~\eqref{eq:stressletCoupling}).

The velocity  matches that of an isolated particle when $d\gg 1$, and the corrections introduced by the particles' interaction scale as $1/d^2$, as a result of the dominant phoretic repulsion (as for case A): all three methods are able to capture that property (see figure~\ref{fig:caseBResults}b,d, inset). Similarly, the second moment of surface concentration decreases as $1/d^3$ (figure~\ref{fig:caseBResults}c).
As $d$ is reduced, the combined effects of strong phoretic repulsion and hydrodynamic coupling (including the repulsion by the active stresslet) slow down and may even eventually reverse the swimming direction of particle 2 (figure~\ref{fig:caseBResults}d).
Both our FCM  solution and the FFA prediction  show a qualitative agreement with the full solution (BSC) and predict the increase in velocity for the leading particle, while the trailing particle is slowed down. However, they  fail to predict the reversal of particle 2's velocity observed in the full solution, although DFCM exhibits an appreciable improvement over FFA in the near field. A possible reason for this may be found in a dominant role of the lubrication layer separating the particles which is not well resolved in either approximation.

\subsection{Janus particles - asymmetric configuration}
Case B was still highly symmetric and further considered only uniform mobility which is known to affect the hydrodynamic signature of the particle significantly~\citep{LaugaMichelin2016}. In our third and final configuration, Case C (figure~\ref{fig:ValidationCasesForPublicationCaseACaseBCaseC}c), we consider a more generic interaction of two identical Janus particles with non-uniform mobility ($\alpha^F_n=0$, $\alpha^B_n=1$, $M^F_n=0$, $M^B_n=1$)  positioned at an angle $\pi/4$ relative to {the $x$-axis}. 
Surface mobility results from the differential short-range interaction of solute and solvent molecules with the particle surface and, as such, is an intrinsic property of the particle's surface coating and may thus differ between the two caps of a Janus particle. 
For these particles, when isolated, the non-dimensional self-propulsion velocity is given by $\boldsymbol{U}^{\infty}=\frac{1}{8}\boldsymbol{e}_x$~\citep{GolestanianLiverpoolAdjari2007}. The convenient bispherical coordinate approach is not usable in this non-axisymmetric setting, and although an extension to generic interactions of Janus particles is possible using full bispherical harmonics~\citep{SharifiMood2016}, it is sufficiently complex that direct numerical simulations using BEM proves in general more convenient, although the discontinuity of the mobility at the equator may introduce numerical errors, due to the singularity of the surface concentration gradient for a Janus particle~\citep{MichelinLauga2014}. In the following, we therefore compare our DFCM predictions with the solution obtained using BEM and the prediction of the far-field analysis (FFA).

\begin{figure}
	\begin{center}
    \includegraphics[width=0.50\textwidth]{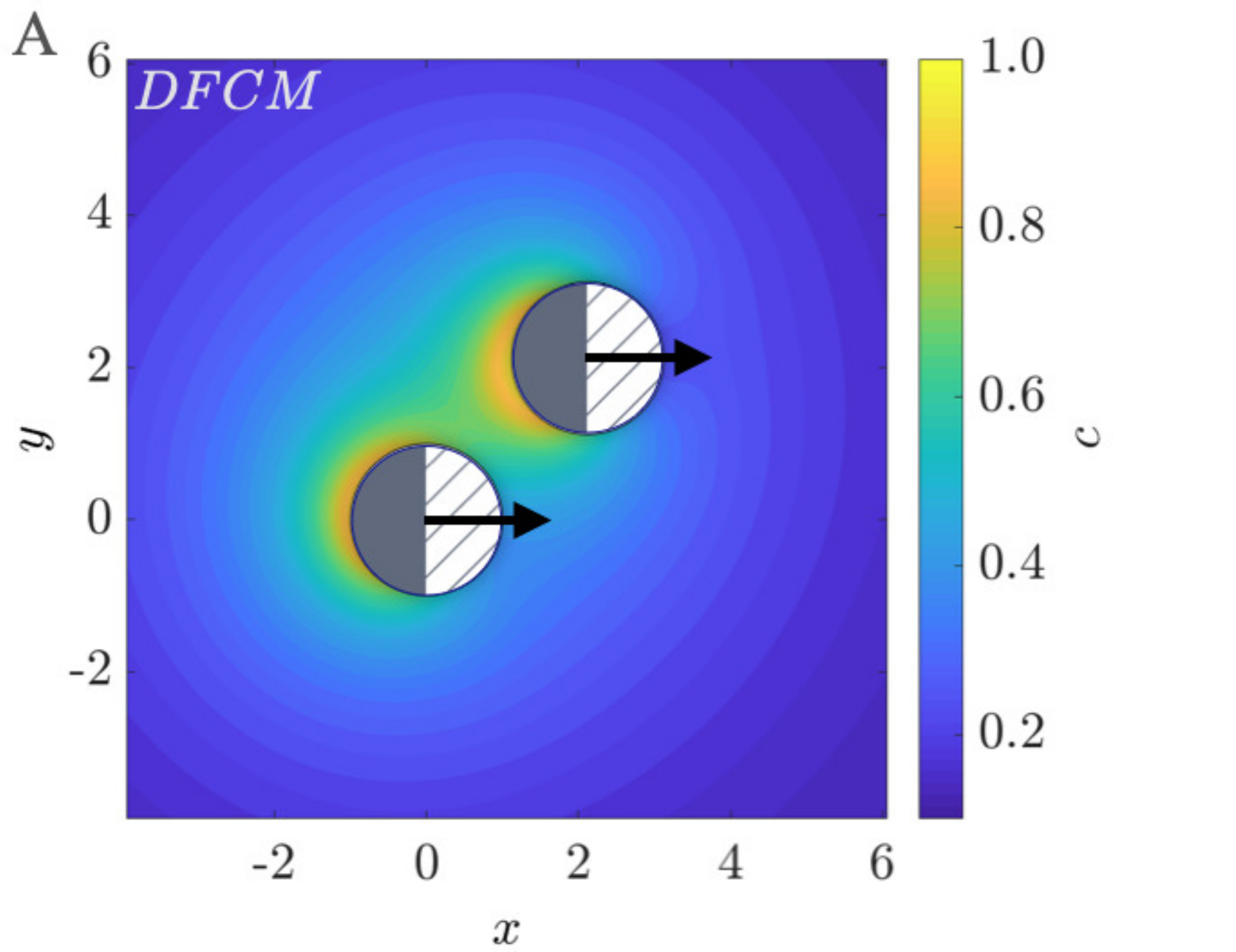}
    \includegraphics[width=0.49\textwidth]{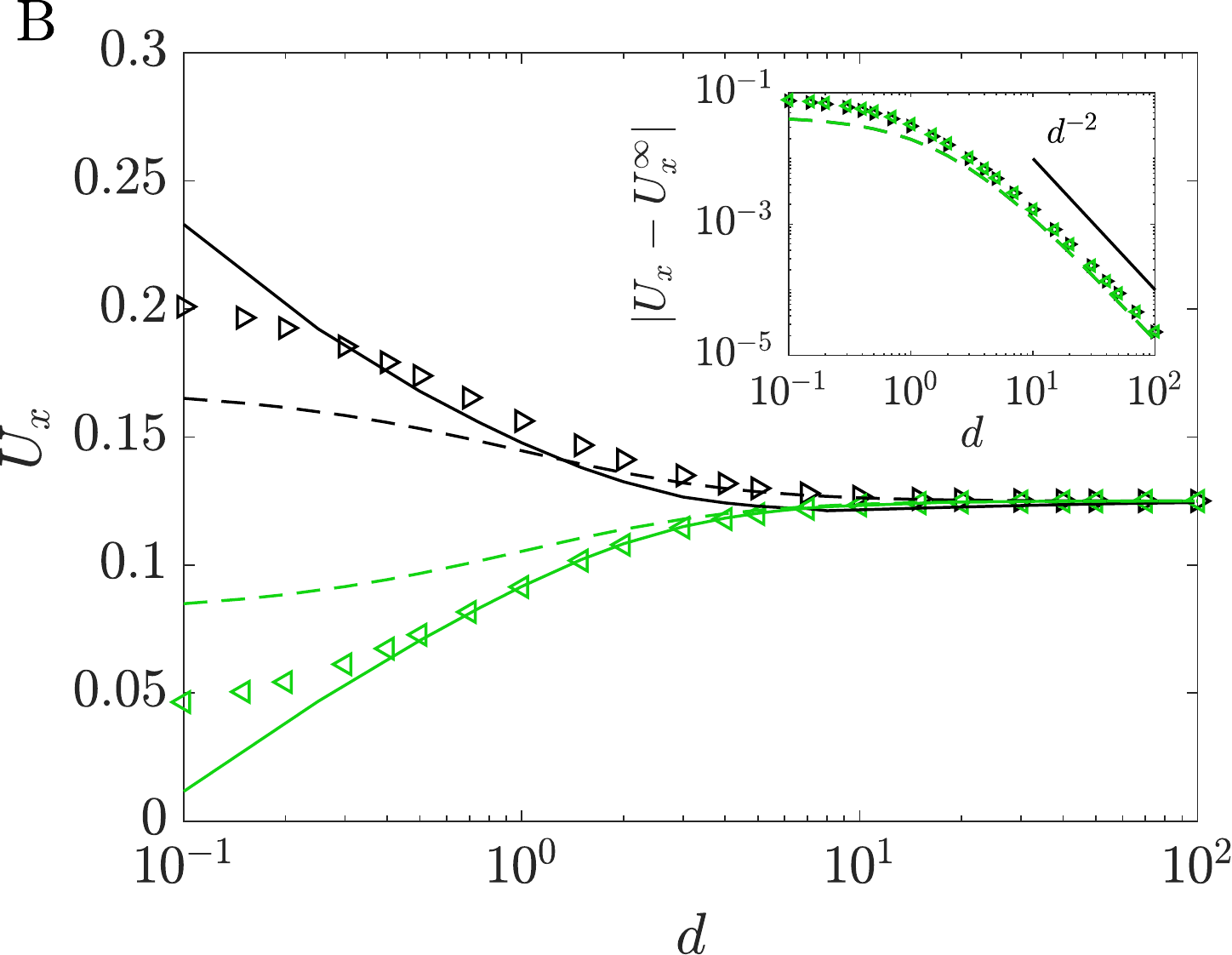}
    \includegraphics[width=0.49\textwidth]{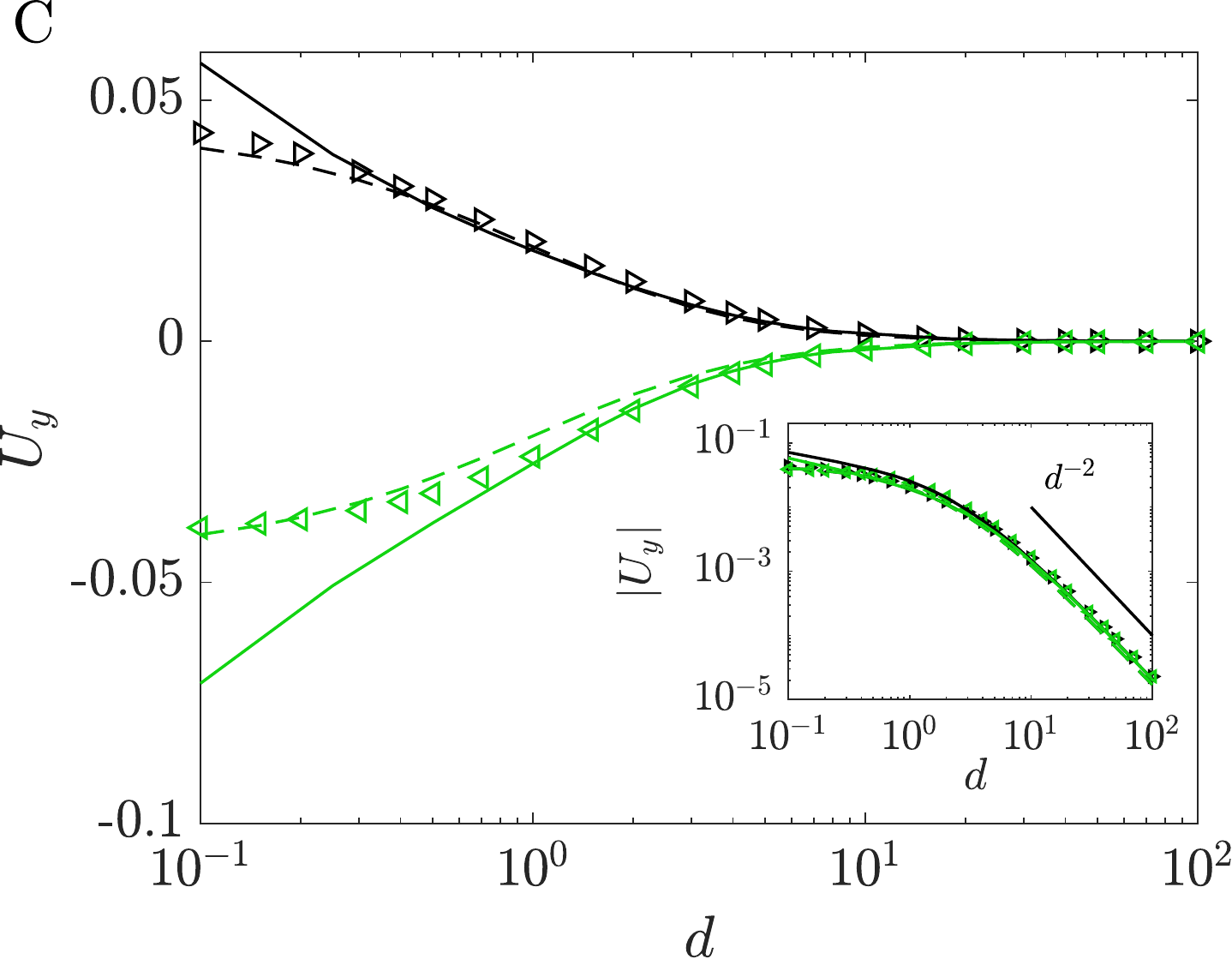}
    \includegraphics[width=0.49\textwidth]{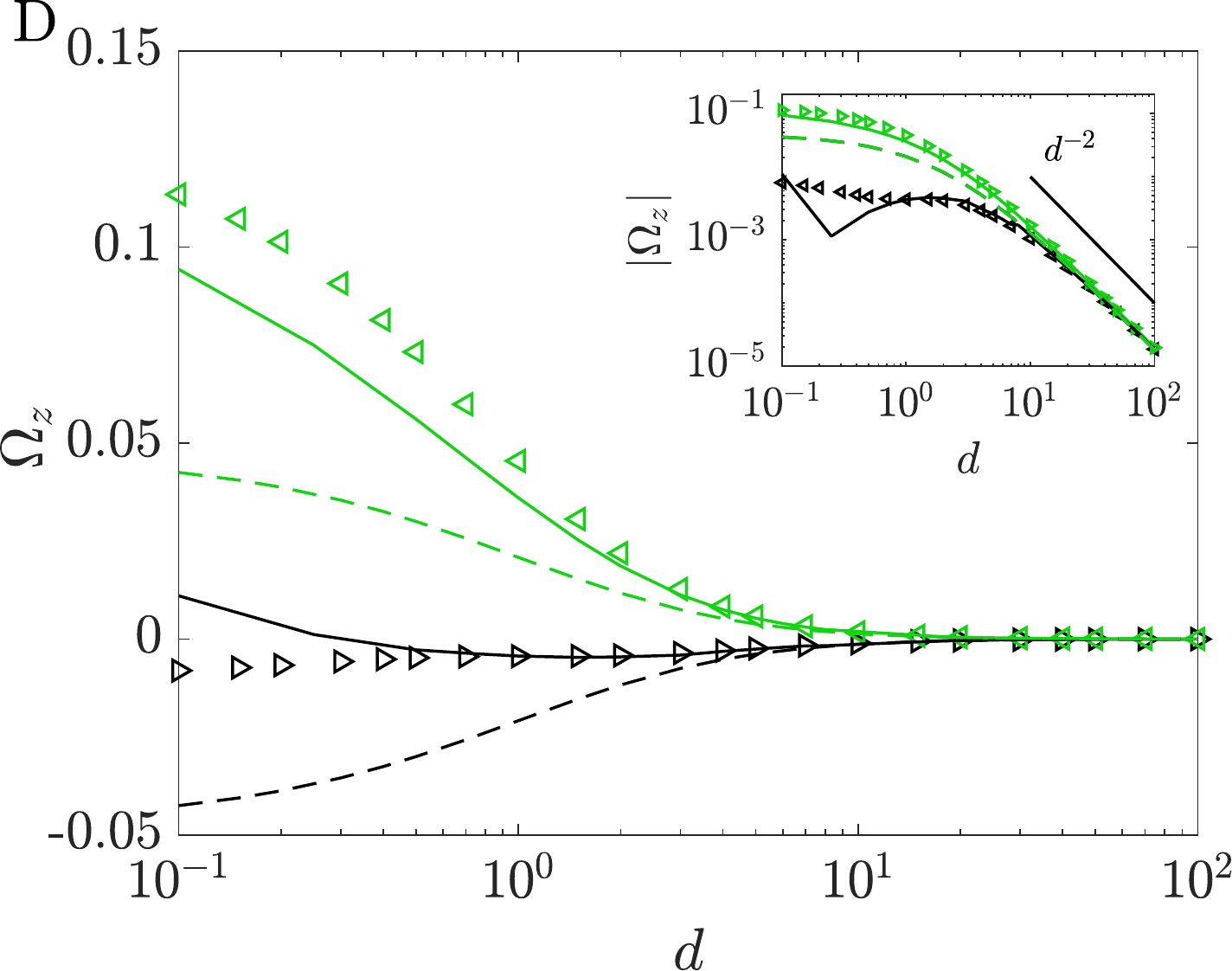}
    \end{center}
    \caption{Case C: a) DFCM concentration  field for $d=1$, b) velocity $U_x$,  c) velocity $U_y$, d) angular velocity $\Omega_z$. The black lines (and markers) correspond to particle 1 and the light green ones to particle 2. The triangle markers correspond to DFCM, the solid lines correspond to BEM, while the dashed lines to FFA. The inset shows the absolute values in logarithmic scale and the corresponding decay.}
    \label{fig:caseCResults}
\end{figure}

The asymmetric concentration field obtained with DFCM for that configuration when $d=1$ is shown on figure~\ref{fig:caseCResults}(a). Besides their intrinsic self-propulsion along $+\boldsymbol{e}_x$ due to their self-generated surface chemical polarity, the accumulation of solute in the confined space between the particles introduces a phoretic repulsion along their line of centers (as for case~B), leading to an enhancement (resp. reduction) of both components of the velocity ($U_x$ and $U_y$) for particle 1 (resp. particle 2). This behaviour is well-captured by all three methods (figure~\ref{fig:caseCResults}b-c). Additionnally, in the present configuration (case C), the mobility is non uniform: specifically here, we consider the case where the surface mobility of the front hemisphere is zero, so that only the back hemisphere generates a phoretic slip. As a result of the arrangement of the particles, the dominant slip along the surface of particle~1 (resp. particle~2) is therefore counter-clockwise (resp. clockwise) leading to a negative (resp. positive) rotation velocity $\Omega_z$ for that particle. This rotation rate is proportional to the polarity, and therefore decays as $1/d^{2}$ in the far field. These intuitive trends are confirmed by the results of all three methods on figure~\ref{fig:caseCResults}(b-d).

As for case B, when the interparticle distance $d$ is reduced, these effects become more pronounced and the results obtained with DFCM for the translation velocity are in that regard slightly better than the predictions of FFA.
 However, FFA predicts a symmetric evolution of $\Omega_z$ with distance, while BEM, the most accurate solution, shows that particle 1 rotates slower than particle 2 for $d<10$, and changes direction in the near field $d<0.2$.  DFCM is able to capture this nontrivial and asymmetric  evolution of the rotation velocity, but fails to capture the direction reversal of particle 1; as for case B, this may stem from the inability of DFCM to  resolve correctly the lubrication flows within the thin fluid gap between the particles.\\
  
Nevertheless, over all three cases considered and in particular in the most generic setting of Janus particles with non-uniform mobility in non-axisymmetric settings, our results show the importance of the proper resolution of higher order hydro-chemical multipolar signatures (e.g. induced polarities and rigidity stresslets) in order to capture accurately  non-trivial feature of the hydro-chemical interactions between particles. DFCM may not be able to resolve the details of the chemical and hydrodynamic fields in the gap between the surface of the particles when they are close to each other (e.g. $d\lesssim 0.5$) as it does not actually represent the exact position of the surface. Yet, this new numerical approach offers significant improvements in capturing such complex effects both qualitatively and quantitatively in comparison with simpler analytical or numerical models, while providing a significant reduction in complexity in comparison with detailed numerical simulations such as BEM, opening significant opportunities for the numerical analysis of larger number of particles and suspension dynamics.

\section{Discussion} \label{sec:conclusions}
In this work, we presented a generalization called Diffusiophoretic Force Coupling Method (DFCM) of the approach of the hydrodynamic FCM in order to compute hydro-chemical interactions within reactive suspensions of Janus particles with non-uniform surface activity and mobility. 
 Following the standard hydrodynamic FCM, we rely on a truncated regularized multipolar expansion at the dipole level to solve the Laplace problem for the reactant concentration field, and its moments at the particle  surface. While the monopole is directly obtained from the prescribed  fluxes on the swimmer surface, the dipole is found iteratively by  accounting for the effect of other particles on their polarity.
  Instead of using surface operators, which are difficult to handle on Eulerian grids, our method relies on spectrally convergent weighted volume averages to compute successive concentration moments. Unlike standard FCM, the averaging envelopes are non Gaussian as their weight is shifted toward the particle's surface and thus differ from the Gaussian spreading envelopes associated with each singularity. 
  The first two moments of concentration around the particle are directly related to the intrinsic phoretic velocity and rotation  of the particles (i.e. those obtained for an isolated particle experiencing the same hydrodynamic surface slip in an unbounded domain) but also to the singularities characterizing their hydrodynamic signatures, i.e. an intrinsic active stresslet and a potential dipole. These multipoles are then used as inputs for the solution of the hydrodynamic (swimming) problem, solved using the  existing hydrodynamic FCM framework to obtain the total particle velocities.

Even though our approximate method does not resolve the particle surface exactly (and is as such unable to capture lubrication or strong confinement effects), its predictions for the dynamics of two particles compare well with analytical or accurate numerical solutions for distances larger than half a radius ($d\gtrsim 0.5$), which is relevant for dilute and semi-dilute suspensions.
Most importantly, in all the results presented above, DFCM provides significant improvements over far-field models that neglect mutually-induced polarities and rigidity stresslets. Our case study has shown the importance of properly resolving these dipolar singularities to capture non-trivial hydro-chemical interactions between particles. 
 
Although the present work purposely focuses on the presentation of the framework and detailed validation on pairwise interactions of phoretic particles, our diffusio-phoretic framework readily generalizes to $N$ particles.
A remarkable feature of FCM is that the spreading and averaging operations are volume-based and independent of the Stokes and Laplace solvers. 
Instead of using Green's functions for specific geometries, the reactant concentration $c$ and fluid velocity  $\boldsymbol{u}$ can be solved for with any numerical method  (e.g. finite volume, spectral methods) on an arbitrary domain where the FCM spreading   and averaging  operations are performed on the fixed computational grid \citep{MaxeyPatel2001,LiuAllKarniadakis2009,YeoMaxey2010}. As shown in previous work \citep{DelmotteAllCliment2015}, the corresponding cost scales linearly with the particle number $O(N)$, while Green's function-based methods, such as  Stokesian Dynamics \citep{BradyBossis1988} and the method of reflections \citep{VarmaMichelin2019}, are restricted to simple geometries and require sophisticated techniques to achieve similar performances  instead of their intrisic quadratic scaling $O(N^2)$ \citep{LiangGreengardJCP2013,FioreSwan2019,Yan2020}.
In addition to improving far-field models, our method  therefore offers a scalable framework for large scale simulations of reactive particles. 
We will use these capacities to study their collective motion  and characterize their macroscopic rheological response. 

Despite its specific focus on the modelling of hydrochemical interactions within phoretic suspensions, the present analysis demonstrates how the fundamental idea of the original Force Coupling Method can be extended and applied to other fields of physics. In such an approach the elliptic Stokes equations are solved over the entire domain (instead of the multiply-connected fluid domain outside the particles) by introducing regularized forcings whose support is calibrated to account for the particle finite size and whose intensity is determined to account for a weak form of the boundary condition. For the chemical diffusion problem considered here, this amounts to (i) replacing a Laplace problem by a Poisson equation, (ii) calibrating the support of the spreading operators to match benchmark properties for a single particle and (iii) determining the forcing intensity by projecting the Neumann-type boundary condition on the particle surface onto a localized support function of appropriate shape (e.g. Gaussian or annular). This approach can readily be adapted for solving diffusion problems with more general (Dirichlet or mixed) boundary conditions, as encountered for more detailed chemical activity of reactive particles~\citep{MichelinLauga2014,TatuleaCodreanLauga2018} or in bubble growth/dissolution problems~\citep{MichelinAllLauga2018}, but also to other physical phenomena driven by elliptic equations, such as electromagnetic interactions of particles~\citep{KeavenyMaxey2019}.

\section*{Acknowledgments}
This work was supported by the European Research Council (ERC) under the European Union's Horizon 2020 research and innovation program (Grant Agreement No. 714027 to S.M.).

\appendix

\section{Determining the source intensities}\label{app:intensities}
We consider here a single active particle bounded by a surface $S$. The concentration field outside $S$ (in the fluid) satisfies Laplace's equation, and its value anywhere in the fluid domain can therefore be obtained in terms of its value and normal flux on $S$ as 
\begin{equation}
    c{(\boldsymbol{r})} = \frac{1}{4\pi} \int_S \left[
    c{(\boldsymbol{s})} \boldsymbol{n} \cdot \frac{(\boldsymbol{r}-\boldsymbol{s})}{|\boldsymbol{r}-\boldsymbol{s}|^3} + \left( -\frac{\partial c{(\boldsymbol{s})}}{\partial n} \right) \frac{1}{|\boldsymbol{r}-\boldsymbol{s}|} \right] \mathrm{d}S.
    \label{eq:integralRepresentationOfLaplaceEquation}
\end{equation}
where $\boldsymbol{s}=a\boldsymbol{n}$ {and $\boldsymbol{r}$ are measured from the center of the particle}. Far from the particle (i.e. $|\boldsymbol{r}|\gg |\boldsymbol{s}|$), and using the following Taylor expansion for $|\boldsymbol{r}-\boldsymbol{s}|^{-n}$,
\begin{equation}
    \frac{1}{|\boldsymbol{r}-\boldsymbol{s}|^n} \approx \frac{1}{r^n} \left[ 1 + n \left(\frac{\boldsymbol{s} \cdot \boldsymbol{r}}{r^2}\right) + n\left(\frac{n}{2}+1\right) \left(\frac{\boldsymbol{s} \cdot \boldsymbol{r}}{r^2}\right)^2 + ...\right],
    \label{eq:approximation1overR}
\end{equation}
the concentration field can be expanded in terms of a series of singular multipoles, namely a  monopole of intensity $q^M$, a dipole of intensity $\boldsymbol{q}^D$, (and up to the desired order of approximation): 
\begin{equation}
    c{(\boldsymbol{r})} = \frac{q^M}{4\pi r} +
    \frac{\boldsymbol{q}^D \cdot \boldsymbol{r}}{4\pi r^3} + \hdots
    \label{eq:concentrationFieldApproximation}
\end{equation}
where the intensities are obtained as:
\begin{equation}
    q^M = \int_{S} \left( -\frac{\partial c{(\boldsymbol{s})}}{\partial n} \right) \mathrm{d}S,
\end{equation}
\begin{equation}
    \boldsymbol{q}^D =
    a \int_{S} \left( -\frac{\partial c{(\boldsymbol{s})}}{\partial n} \boldsymbol{n} \right) \mathrm{d}S
    + \int_{S} c{(\boldsymbol{s})} \boldsymbol{n} \ \mathrm{d}S.
    \label{eq:dipoleExpression}
\end{equation}
Substitution of the boundary condition Eq.~\eqref{eq:boundaryConditionDiffusionSphereN} leads to the result in Eq.~\eqref{eq:monopoledipoleIntensity_FINAL}.

\section{Intrinsic phoretic velocities and stresslet}\label{app:velocitycoupling}
The intrinsic phoretic velocity of a particle (i.e. its swimming speed in the absence of any hydrodynamic interactions or outer flow) is defined in Eq.~\eqref{eq:translationalAndRotationalVelocitiesCoupling}. Using the slip velocity definition in Eq.~\eqref{eq:boundaryConditionHydrodynamicsSphereN} and the mobility distribution as in Eq.~\eqref{eq:trueJanus_1_2_mobiAlteExpr}, we obtain:
\begin{equation}
    \boldsymbol{U}^a_n \ = \ -\langle \boldsymbol{u}_s \rangle_n \ = -\overline{M}_n \langle \nabla_{\parallel} c \rangle_n - M_n^* \langle \mathrm{sign}(\boldsymbol{p} \cdot \boldsymbol{n}) \ \nabla_{\parallel} c \rangle_n.
    \label{eq:photeticparticleVelocityA2}
\end{equation}
Integrating by parts the surface averaging operators we arrive to:
\begin{equation}
    \boldsymbol{U}^a_n = -\frac{2\overline{M}_n}{a} \langle c\boldsymbol{n} \rangle_n + \frac{M_n^* \boldsymbol{p}_n}{a} \langle c  \rangle_n^\textrm{eq}- \frac{M_n^*}{a} \Big( \langle c\boldsymbol{n} \rangle_n^+ - \langle c\boldsymbol{n} \rangle_n^- \Big),
    \label{eq:photeticparticleVelocityB2}
\end{equation}
where the operators $\langle...\rangle_n^\pm$ refer to the mean value over the front and back caps of particle $n$, respectively, and $\langle\hdots\rangle_n^\textrm{eq}$ is the line average over the equator of particle $n$. To compute these particular averages, we expand the surface concentration $c(\boldsymbol{n})$ in terms of its surface moments and truncate the expansion to the first three terms:
\begin{equation}
    c(\boldsymbol{n}) = \langle c \rangle_n + 3 \langle c\boldsymbol{n} \rangle_n \cdot \ \boldsymbol{n} + \frac{15}{2} \langle c(\boldsymbol{n}\boldsymbol{n}-\mathbf{I}/3) \rangle_n \ : \boldsymbol{n}\boldsymbol{n}.
    \label{eq:cAsFunctionOfMomentsOfC}
\end{equation}
 Substitution in Eq.~\eqref{eq:photeticparticleVelocityA2} then finally provides
\begin{equation}
    \boldsymbol{U}^a_n = -\frac{2\overline{M}_n}{a} \langle c\boldsymbol{n} \rangle_n - \frac{15 M^*_n}{8 a} \langle c(\boldsymbol{n}\boldsymbol{n}-\mathbf{I}/3) \rangle_n \ : \big[ \boldsymbol{p}_n\mathbf{I} + (\boldsymbol{p}_n\mathbf{I})^{\mathrm{T}_{12}} + \boldsymbol{p}_n\boldsymbol{p}_n\boldsymbol{p}_n \big],
    \label{eq:photeticparticleVelocityC}
\end{equation}
which can be simplified into Eq.~\eqref{eq:photeticparticleVelocityE} using the symmetry and traceless property of $\boldsymbol{n}\boldsymbol{n}-\mathbf{I}/3$.

 Following a similar procedure, the intrinsic phoretic angular velocity can be expanded from Eqs.~\eqref{eq:boundaryConditionHydrodynamicsSphereN}, \eqref{eq:translationalAndRotationalVelocitiesCoupling} and  \eqref{eq:trueJanus_1_2_mobiAlteExpr}  as
\begin{equation}
    \boldsymbol{\Omega}^a_n = -\frac{3}{2a} \langle \boldsymbol{n} \times M \nabla_{\parallel} c \rangle_n =
    -\frac{3}{2a} \overline{M}_n \langle \boldsymbol{n} \times \nabla_{\parallel} c \rangle_n
    -\frac{3}{2a} M^*_n \langle \mathrm{sign}(\boldsymbol{p} \cdot \boldsymbol{n}) \ \boldsymbol{n} \times \nabla_{\parallel} c \rangle_n,
    \label{eq:driftRotationalVelocityGeneralExpressionBB}
\end{equation}
and after integration by parts simplifies to:
\begin{equation}
    \boldsymbol{\Omega}^a_n = -\frac{3M^*_n}{2 a^2}\left(\boldsymbol{p}_n\times \langle c \boldsymbol{n}\rangle^\textrm{eq}_n\right).
\end{equation}
Substitution of Eq.~\eqref{eq:cAsFunctionOfMomentsOfC} provides the desired expression, Eq.~\eqref{eq:driftRotationalVelocityFCMExpression}.\\

The same method can also be applied to determine the intrinsic phoretic stresslet $\mathbf{S}^a_n$. From its definition in Eq.~\eqref{eq:stressletCoupling} and using Eqs.~\eqref{eq:boundaryConditionHydrodynamicsSphereN} and \eqref{eq:trueJanus_1_2_mobiAlteExpr}, we obtain: 
\begin{equation}
    \mathbf{S}^a_n = -10\pi a^2 \overline{M}_n \langle (\boldsymbol{n} \nabla_{\parallel} c + (\nabla_{\parallel} c) \ \boldsymbol{n} ) \rangle_n -10\pi a^2 M^*_n \langle \textrm{sign}(\boldsymbol{p} \cdot \boldsymbol{n}) (\boldsymbol{n} \nabla_{\parallel} c + (\nabla_{\parallel} c) \ \boldsymbol{n} ) \rangle_n
    \label{eq:photeticparticleStressletB2}
\end{equation}
Integrating by parts the surface averaging operators provides
\begin{align}
    \mathbf{S}^a_n = &-60 \pi a \overline{M}_n \langle c(\boldsymbol{n}\boldsymbol{n}-\mathbf{I}/3) \rangle_n\nonumber\\
    &+ 10 \pi a M^*_n \Big[ \langle c\boldsymbol{n}\rangle_n^\textrm{eq}  \boldsymbol{p}_n + \boldsymbol{p}_n \langle c\boldsymbol{n}\rangle_n^\textrm{eq}  
    - 3 \Big(\langle c(\boldsymbol{n}\boldsymbol{n}-\mathbf{I}/3) \rangle_n^+ - \langle c(\boldsymbol{n}\boldsymbol{n}-\mathbf{I}/3) \rangle_n^-\Big)\Big]
    \label{eq:photeticparticleStressletB3}
\end{align}
Subsitution of Eq.~\eqref{eq:cAsFunctionOfMomentsOfC} provides finally
\begin{align}
    \mathbf{S}^a_n = -60 \pi a \overline{M}_n \langle c(\boldsymbol{n}\boldsymbol{n}-\mathbf{I}/3) \rangle_n
    + \frac{15}{2} \pi a M^*_n \Big[ \Big(\langle c\boldsymbol{n}\rangle_n\cdot \boldsymbol{p}_n\Big)(\mathbf{I}-\boldsymbol{p}_n\boldsymbol{p}_n)-\langle c\boldsymbol{n}\rangle_n\boldsymbol{p}_n     -\boldsymbol{p}_n\langle c\boldsymbol{n}\rangle_n  \Big].
    \label{eq:photeticparticleStressletC}
\end{align}


\end{document}